\documentclass[12pt,preprint]{aastex}

\usepackage{CJK}
\usepackage{natbib}
\usepackage{graphicx}
\usepackage{rotating}

\begin{document}
\begin{CJK}{UTF8}{gbsn}

\title{Two Beyond-Primitive Extrasolar Planetesimals}

\author{S. Xu(许\CJKfamily{bsmi}偲\CJKfamily{gbsn}艺)\altaffilmark{a}, M. Jura\altaffilmark{a}, B. Klein\altaffilmark{a}, D. Koester\altaffilmark{b}, B. Zuckerman\altaffilmark{a}}
\altaffiltext{a}{Department of Physics and Astronomy, University of California, Los Angeles CA 90095-1562; sxu@astro.ucla.edu, jura@astro.ucla.edu, kleinb@astro.ucla.edu, ben@astro.ucla.edu}
\altaffiltext{b}{Institut fur Theoretische Physik und Astrophysik, University of Kiel, 24098 Kiel, Germany; koester@astrophysik.uni-kiel.de}

\begin{abstract}
Using the Cosmic Origins Spectrograph onboard the {\it Hubble Space Telescope}, we have obtained high-resolution ultraviolet observations of GD 362 and PG 1225-079, two helium-dominated, externally-polluted white dwarfs. We determined or placed useful upper limits on the abundances of two key volatile elements, carbon and sulfur, in both stars; we also constrained the zinc abundance in PG 1225-079. In combination with previous optical data, we find strong evidence that each of these two white dwarfs has accreted a parent body that has evolved beyond primitive nebular condensation. The planetesimal accreted onto GD 362 had a bulk composition roughly similar to that of a mesosiderite meteorite based on a reduced chi-squared comparison with solar system objects; however, additional material is required to fully reproduce the observed mid-infrared spectrum for GD 362. No single meteorite can reproduce the unique abundance pattern observed in PG 1225-079; the best fit model requires a blend of ureilite and mesosiderite material. From a compiled sample of 9 well-studied polluted white dwarfs, we find evidence for both primitive planetesimals, which are a direct product from nebular condensation, as well as beyond-primitive planetesimals, whose final compositions were mainly determined by post-nebular processing.
\end{abstract}


\section{INTRODUCTION}

Planetesimals are building blocks of planets and their formation is a key step towards planet formation. How do planetesimals form? What determines their bulk composition? To answer these questions, we start by examining our own solar system.

The overall configuration of the solar system is that volatile-depleted, dry rocky objects are ubiquitous relatively close to the Sun while volatile-rich, icy objects are found beyond the snow line. This correlation between the volatile fraction and heliocentric distance can be explained by primitive nebular condensation: refractory elements condensed closer to the Sun while volatile elements can only be incorporated into the planetesimals where the temperature is low enough. Many solar system objects have experienced some additional processing that changed their initial compositions. For example, it has been argued that a collision between a large asteroid and proto-Mercury stripped off most of Mercury's silicate mantle, leaving it $\sim$70\% iron by mass \citep{Benz1988}. Also, the ``late veneer" has delivered a large amount of water and volatiles onto Earth \citep{Chyba1990}. Post-nebular processing, such as collisions, melting and differentiation, is important in redistributing the elements among solar system objects.

Currently, the best way to measure the elemental compositions of planetesimals in the solar system is from meteorites, which are fragments from collisions among asteroids. Following \citet{ONeillPlame2008}, we classify all meteorites into two categories in this paper. (i) ``Chondritic" is used to refer to chondrites, which are a direct product of nebular processing. Objects in this category are described as ``primitive" planetesimals. (ii) ``Non-chondritic" objects consist of achondrites, stoney-iron meteorites and iron meteorites. Examples of their parent bodies include the Moon, Mars or asteroids that have experienced various amounts of post-nebular processing. Planetesimals in this category are considered to be ``beyond-primitive". 

What about planetesimal formation in extrasolar planetary systems? High-resolution, high-sensitivity spectroscopic observations of externally-polluted white dwarfs are a powerful tool for determining the bulk elemental compositions of extrasolar planetesimals \citep{Jura2013}. Calculations show that minor planets can survive the red giant stage of a star and persist into the white dwarf phase with most of their internal water and volatiles intact \citep{Jura2008, JuraXu2010}. Orbital perturbations from one or multiple planets can cause these planetesimals to stray into the tidal radius of the white dwarf and get tidally disrupted \citep{DebesSigurdsson2002, Bonsor2011, Debes2012a}, sometimes producing a dust disk that emits mostly in the infrared \citep{Jura2003, Kilic2006b, VonHippel2007, Farihi2009, XuJura2012}. Eventually, all this planetary debris is accreted onto the central white dwarf and pollutes its otherwise pure hydrogen or helium atmosphere. 

The first comprehensive abundance measurement of an externally-polluted white dwarf was performed by \citet{Zuckerman2007}, who identified 15 elements heavier than helium in the atmosphere of GD 362, including Mg, Si and Fe, which are often called the ``common elements" \citep{Larimer1988}. The disrupted object had a minimum mass $\sim$10$^{22}$ g, which is comparable to that of a massive solar system asteroid. Three years later, the abundances of eight heavy elements were determined in the atmosphere of GD 40, including all the major rock-forming elements -- O, Mg, Si and Fe \citep{Klein2010}. Now there are many more high-resolution optical spectroscopic studies of externally-polluted white dwarfs [e.g., \citet{Klein2011,Melis2011, Zuckerman2011, Farihi2011a,Dufour2012, Vennes2010, Vennes2011a}].

However, optical spectroscopy of externally-polluted white dwarfs typically does not enable sensitive detection of highly-volatile elements, such as carbon, nitrogen and sulfur, which are key to understanding the thermal history of the system. Ultraviolet spectroscopy is complimentary to optical observations in determination of volatile abundances.

To-date, there are four white dwarfs with both published high-resolution optical and ultraviolet measurements\footnote{The four white dwarfs are: GD 61 \citep{Desharnais2008, Farihi2011a}; GD 40, G241-6 \citep{Klein2010, Klein2011, Zuckerman2010, Jura2012} and WD 1929+012 \citep{Vennes2010, Vennes2011a, Melis2011, Gaensicke2012}.}; we are beginning to accumulate an atlas of the compositions of extrasolar planetesimals. To zeroth order, we find that they are strikingly similar to meteorites in the solar system: (i) O, Mg, Si and Fe are always dominant and their sum is more than 85\% of the accreted mass; (2) volatile elements, especially C, are typically depleted by more than a factor of 10 compared to solar abundances\footnote{Very recently, \citet{Koester2012} reported several white dwarfs with solar carbon-to-silicon ratio. However, the source of this pollution is unclear and more analysis is forthcoming. }.

In this paper, we report ultraviolet spectroscopic observations of GD 362 and PG 1225-079 with the Cosmic Origins Spectrograph (COS) onboard the {\it Hubble Space Telescope} ({\it HST}), complimentary to previous optical studies from the Keck High Resolution Echelle Spectrometer (HIRES) \citep{Zuckerman2007, Klein2011}. PG 1225-079 has been observed with the low-resolution International Ultraviolet Explorer (IUE) \citep{Wolff2002}; there is no previous ultraviolet spectroscopy for GD 362. The rest of the paper is organized as follows. Data reduction is summarized in section 2 and atmospheric abundance determinations are reported in section 3. In section 4, we used a reduced chi-squared analysis to look for solar system analogs to the accreted parent bodies. The formation mechanisms of extrasolar planetesimals are assessed in section 5 and conclusions are given in section 6. In Appendix A, we report the {\it Herschel} Photodetecting Array Camera and Spectrometer (PACS) observation of GD 362. In Appendix B, we extend the reduced chi-squared analysis to two additional externally-polluted helium white dwarfs with both high-resolution optical and ultraviolet observations.

\section{OBSERVATIONS AND DATA REDUCTION}

GD 362 and PG 1225-079 were observed during {\it HST}/COS Cycle 18 under program 12290. These two white dwarfs are too cool to be observed effectively with the G130M grating centering around 1300 {\AA}, as was employed by \citet{Jura2012} and \citet{Gaensicke2012} for other hotter white dwarfs. Instead, the G185M grating was used with a central wavelength of 1921 {\AA} and wavelength coverage of 1800 -- 1840 {\AA}, 1903 -- 1940 {\AA} and 2008 -- 2044 {\AA}. The spectral resolution was $\sim$18,000. Total exposure times were 7411 and 1805 sec for GD 362 and PG 1225-079, respectively.

The raw data were processed using the standard pipeline CALCOS 2.13.6. The fluxes at 2030 {\AA} are 2.9 $\times$ 10$^{-15}$ erg s$^{-1}$ cm$^{-2}$ {\AA}$^{-1}$ and 1.5 $\times$ 10$^{-14}$ erg s$^{-1}$ cm$^{-2}$ {\AA}$^{-1}$ for GD 362 and PG 1225-079, respectively, in approximate agreement with broadband NUV fluxes from the {\it GALEX} satellite. The signal-to-noise ratio (SNR) in the original un-smoothed spectrum was 6 for PG 1225-079 and 4 for GD 362.

Following previous data reduction procedures \citep{Klein2010, Klein2011, Jura2012}, for PG 1225-079, equivalent widths (EWs) of each spectral line were measured in the un-smoothed spectra by fitting a Voigt profile with three different nearby continuum intervals in IRAF. The EW uncertainty is calculated by adding the standard deviation of the three EWs and the average uncertainty from the profile fitting in quadrature. The EW upper limit is obtained by artificially inserting a spectral line with different abundance into the model and comparing with the data. We adopt a different method to measure the EW for C I 1930.9 {\AA} in GD 362, as described in section 3.1. The measured values are listed in Tables \ref{Tab: LinesGD} and \ref{Tab: LinesPG} for GD 362 and PG 1225-079, respectively. The average Doppler shift relative to the Sun for PG 1225-079 is 42 $\pm$ 13 km s$^{-1}$, in essential agreement with the value 49 $\pm$ 3 km s$^{-1}$ derived from optical studies \citep{Klein2011}. The large velocity dispersion in the ultraviolet is due to the low SNR of the spectrum and the $\sim$ 15 km s$^{-1}$ uncertainty of COS (COS Instrument Handbook). For GD 362, we marginally detected C I 1930.9 {\AA} and it has a Doppler shift of 48 km s$^{-1}$, in agreement with 49.3 $\pm$ 1.0 km s$^{-1}$ from the optical study \citep{Zuckerman2007}. 

\begin{table}[htpb]
\begin{center}
\caption{Measured Equivalent Widths and Abundance Determinations for GD 362}
\begin{tabular}{lccccccc}
\\
\hline \hline
Ion	& $\lambda$	& E$_{low}$	& EW	& log n(Z)/n(He)	\\
	& (\AA)	& (eV)	& (m\AA)	& \\
\hline
C I	& 1930.905	&	1.26	& 560 $^{+230}_{-158}$ $^a$ & -6.70 $\pm$ 0.30\\
\\
S I	& 1807.311	&	0	& $\lesssim$ 900	 & $\lesssim$ -6.70 	\\
S I	& 1820.341	&	0.049 & $\lesssim$ 710	& $\lesssim$ -6.40	\\
S 	&	&	&	& $\lesssim$ -6.70\\
\hline
\label{Tab: LinesGD}
\end{tabular}
\end{center}
$^a$ This is measured from the model spectra, as described in section 3.1.
\end{table}

\begin{table}[htpb]
\begin{center}
\caption{Measured Equivalent Widths and Abundance Determinations for PG 1225-079}
\begin{tabular}{lcccccc}
\\
\hline \hline
\\
Ion	& $\lambda$	& E$_{low}$	& EW	&  log n(Z)/n(He)	\\
	& (\AA)	& (eV)	& (m\AA)	&\\
\hline
C I	& 1930.905	&	1.26	& 1600 $\pm$ 200	& -7.80 $\pm$ 0.10	\\
\\
S I	& 1807.311	&	0	& $\lesssim$ 170 & $\lesssim$ -9.50	 \\
S I	& 1820.341	&	0.049 & $\lesssim$ 150	& $\lesssim$ -9.30	\\
S 	&	&	&	& $\lesssim$ -9.50 \\
\\
Mg I	& 2026.477$^a$	&	0	& 288 $\pm$ 100$^b$	& $\lesssim$ -7.60	\\
\\
Si II	& 1808.013	&	0	& 936 $\pm$ 109	& -7.44 $\pm$ 0.10 \\
Si II	& 1816.928	&	0.04	& 1232 $\pm$ 145	& -7.46 $\pm$ 0.10	\\
Si	&			&	&	& -7.45 $\pm$ 0.10\\
\\
Fe II	& 1925.987	&	2.52	& 192 $\pm$ 72	& -7.62 $\pm$ 0.28	\\
Fe II	& 2011.347	&	2.58	& 309 $\pm$ 97	& -7.35 $\pm$ 0.24	\\
Fe II	& 2019.429	& 	1.96	&  211 $\pm$ 67	& -7.56 $\pm$ 0.24 \\
Fe II & 2021.402	&	1.67	&  181 $\pm$ 68	& -7.71 $\pm$ 0.27\\
Fe II & 2033.061	&	2.03	&  311 $\pm$ 67	& -7.24 $\pm$ 0.17\\
Fe II	& 2041.345	&	1.964 & 215 $\pm$ 50	& -7.24 $\pm$ 0.18 \\
Fe	&	&	&	&  -7.45 $\pm$ 0.23\\
\\
Zn II	& 2026.136	&	0	& 288 $\pm$ 47$^b$	& $\lesssim$ -11.30	\\
\hline
\label{Tab: LinesPG}
\end{tabular}
\end{center}
$^a$ The atomic parameters for this line are taken from \citet{KelleherPodobedova2008}. \\
$^b$ Mg I 2026.5 {\AA} and Zn II 2026.1 {\AA} are blended and the reported EW is for the entire feature.
\end{table}

\section{ATMOSPHERIC ABUNDANCE DETERMINATIONS}

Because we are most interested in the abundance of an element relative to other heavy elements and these ratios are not strongly dependent upon the stellar temperature and surface gravity \citep{Klein2011}, we only adopt one set of stellar parameters as listed in Table \ref{Tab: Properties} and compute the model spectra following \citet{Koester2010}. Atomic data are mostly taken from the Vienna Atomic Line Database \citep{Kupka1999}. The computed model atmosphere spectra were convolved with the COS NUV line spread function\footnote{http://www.stsci.edu/hst/cos/performance/spectral\_resolution/nuv\_model\_lsf}. The abundance of each element was derived by comparing the EW of each spectral line with the value derived from the model atmosphere, as shown in Figures \ref{Fig: GD_C}-\ref{Fig: PG_Zn} and Tables \ref{Tab: LinesGD} and \ref{Tab: LinesPG}. The final abundances, combining ultraviolet with optical observations, are given in Tables \ref{Tab: AbundanceGD} and \ref{Tab: AbundancePG} for GD 362 and PG 1225-079, respectively. Our results mostly agree with previous reports but have a higher accuracy. For PG 1225-079, we newly derive the abundances of carbon and silicon and have tentative detections of sulfur and zinc. The magnesium abundance is updated while the iron abundance agrees with previous optical results. Because the data are noisier for GD 362, we are only able to crudely constrain the abundance of carbon and sulfur.

\begin{table}[htpb]
\begin{center}
\caption{Adopted Stellar Properties} 
\begin{tabular}{lcccccccc}
\\
\hline \hline
star	& M$_*$	& T	& log g	& D	& log M$_{cvz}$/M$_*$$^a$		& Ref \\
& (M$_{\odot}$)	& (K)		& (cm$^2$ s$^{-1}$)	& (pc)	&	 \\
\hline
GD 362 	     & 0.72	& 10,540	& 8.24	& 51	& -6.71 	& (1) (2)  \\
PG 1225-079 & 0.58	&10,800	& 8.00 	& 26 	& -5.02 	& (3) (4) \\
\hline 
\label{Tab: Properties}
\end{tabular}  
\end{center}
$^a$ Newly-derived mass of the convective zone (see section 4).\\
{\bf References.}{(1) \citet{Kilic2008b}; (2) \citet{Zuckerman2007}; (3) \citet{Klein2011}; (4) \citet{Farihi2005}.}\\
\end{table}

\begin{table}[htpb]
\begin{center}
\caption{Atmospheric Abundances for GD 362\label{Tab: AbundanceGD}}
\begin{tabular}{lllccc}
\\
\hline \hline 
Z	& log n(Z)/n(He)$^a$		& t$_{set}$$^b$	& $\dot{M}$(Z$_i$)$^c$	\\
	&	&	(10$^5$ yr)	& (g s$^{-1})$	\\
\hline 
H	& -1.14 $\pm$ 0.10	& ...	& ...		\\
C$^*$	& -6.70 $\pm$ 0.30	& 2.1	& 2.5 $\times$ 10$^7$	\\
N	& $<$ -4.14		& 2.2	& $<$ 9.0 $\times$ 10$^9$		\\
O	& $<$ -5.14		& 2.2	& $<$ 1.1 $\times$ 10$^9$\\
Na	& -7.79 $\pm$ 0.20	& 2.2	& 3.7 $\times$ 10$^6$		\\
Mg	& -5.98 $\pm$ 0.25	& 2.2	& 2.5 $\times$ 10$^8$	\\
Al	& -6.40 $\pm$ 0.20	& 1.6	& 1.5 $\times$ 10$^8$		\\
Si	& -5.84 $\pm$ 0.30	& 1.2	& 7.2 $\times$ 10$^8$	\\
S$^*$	& $\lesssim$ -6.70$^d$		& 0.79	& $\lesssim$ 1.7 $\times$ 10$^8$		\\
Ca	& -6.24 $\pm$ 0.10 	& 0.99	& 5.1 $\times$ 10$^8$		\\
Sc	& -10.19 $\pm$ 0.30	& 0.93	& 6.8 $\times$ 10$^4$	\\
Ti	& -7.95 $\pm$ 0.10	& 0.94	& 1.2 $\times$ 10$^7$	\\
V	& -8.74 $\pm$ 0.30	& 0.95	& 2.1 $\times$ 10$^6$	\\
Cr	& -7.41 $\pm$ 0.10	& 1.0	& 4.3 $\times$ 10$^7$	\\
Mn	& -7.47 $\pm$ 0.10	& 1.0& 4.0 $\times$ 10$^7$	\\
Fe	& -5.65 $\pm$ 0.10	& 1.1& 2.5 $\times$ 10$^9$	\\
Co	& -8.50 $\pm$ 0.40	& 0.99& 4.1 $\times$ 10$^6$	\\
Ni	& -7.07 $\pm$ 0.15	& 1.0& 1.1 $\times$ 10$^8$	\\
Cu	& -9.20 $\pm$ 0.40	& 0.83& 1.1 $\times$ 10$^6$	\\
Sr	& -10.42 $\pm$ 0.30	& 0.56& 1.3 $\times$ 10$^5$	\\
Total	&	&	& 4.4 $\times$ 10$^9$	\\
\hline
\end{tabular}
\end{center}
$^*$ New measurements from this paper. The rest are from \citet{Zuckerman2007} but we reference abundances relative to He, the dominant element in GD 362's atmosphere, rather than H, as presented in \citet{Zuckerman2007}. Consequently, there is a possible systematic offset up to 0.1 dex in all entries derived from that paper.\\
$^a$ The final abundance of an element combining optical and ultraviolet data.\\
$^b$ Newly-derived settling times in the convective zone (see section 4); they are typically a factor of 2-3 longer than previously-derived values in \citet{Koester2009a}. \\
$^c$ Accretion rates calculated from Equation (1).\\
$^d$ The equality sign corresponds to the red model fit shown in figures.
\end{table}

\begin{table}[htpb]
\begin{center}
\caption{Atmospheric Abundances for PG 1225-079\label{Tab: AbundancePG}}
\begin{tabular}{lccc}
\\
\hline \hline 
Z	& log n(Z)/n(He)	& t$_{set}$	& $\dot{M}$(Z$_i$)	\\
	&	&	(10$^6$ yr)	& (g s$^{-1})$	\\
\hline 
H	& -4.05 $\pm$ 0.10	& ...	& ... 	\\
C$^*$	& -7.80 $\pm$ 0.10	& 5.5	& 3.1 $\times$ 10$^6$	\\
O	& $<$ -5.54	& 	4.5	& $<$ 9.1$\times$ 10$^8$ \\
Na	& $<$ -8.26	& 	4.4& $<$ 2.6 $\times$ 10$^6$ \\
Mg$^*$	&	-7.50 $\pm$ 0.20	& 4.8& 1.4 $\times$ 10$^7$	\\
Al	&	$<$ -7.84	& 3.6	& $<$ 9.5 $\times$ 10$^6$		\\
Si$^*$	&	-7.45 $\pm$ 0.10 & 	3.0	& 3.0 $\times$ 10$^7$	\\
S$^*$	& 	$\lesssim$ -9.50	& 1.7	& $\lesssim$ 5.2 $\times$ 	10$^5$	\\
Ca	&	-8.06	 $\pm$ 0.03 	& 	1.9	& 1.6 $\times$ 10$^7$	\\
Sc	&	-11.29 $\pm$ 0.07	& 	1.8	& 1.1 $\times$ 10$^4$ \\
Ti	&	-9.45 $\pm$ 0.02	&	1.8	& 8.3 $\times$ 10$^5$\\
V	&	-10.41 $\pm$ 0.10	&	1.8	& 9.6 $\times$ 10$^4$\\
Cr	& 	-9.27 $\pm$ 0.06	& 	1.9	& 1.3 $\times$ 10$^6$\\
Mn	& 	-9.79 $\pm$ 0.14	&	2.0	& 4.0 $\times$ 10$^5$ 	\\
Fe	& 	-7.42 $\pm$ 0.07	& 	2.1	& 9.0 $\times$ 10$^7$\\
Ni	& 	-8.76 $\pm$ 0.14	&	2.3	& 4.0 $\times$ 10$^6$	\\
Zn$^*$	&	$\lesssim$ -11.30		& 2.2		& $\lesssim$ 1.3 $\times$ 10$^4$\\
Sr	& 	$<$ -11.65 		& 1.2		& $<$ 1.4 $\times$ 10$^4$\\
Total	&	&	& 1.6 $\times$ 10$^8$ \\
\hline
\end{tabular}
\end{center}
$^*$ New results from this paper. The rest are from \citet{Klein2011}.\\
{\bf Notes.} The columns are defined the same as Table \ref{Tab: AbundanceGD}.
\end{table}

\subsection{Carbon}

\begin{figure}[hp]
\epsscale{1.0}
\plotone{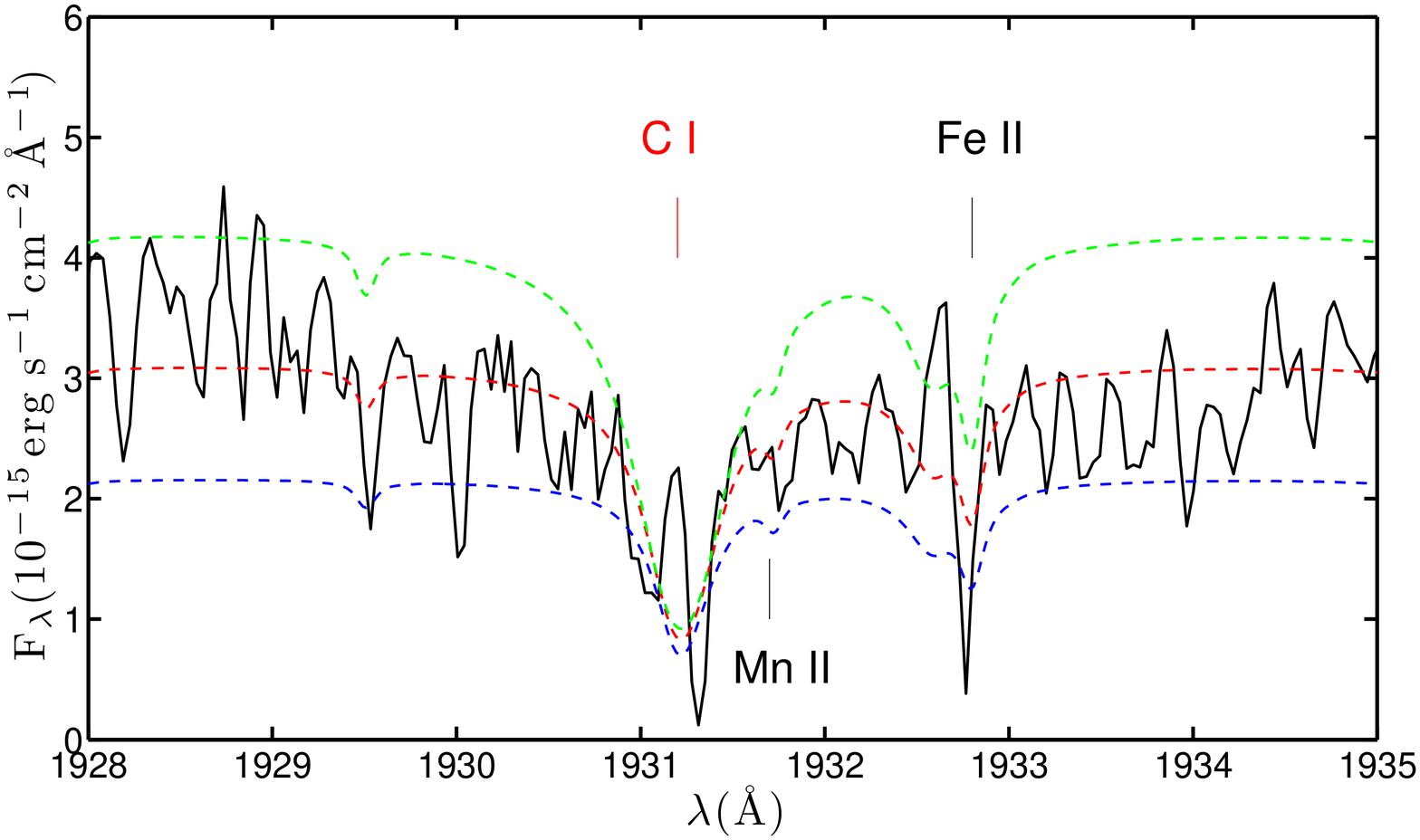}
\caption{{\it HST}/COS spectrum of GD 362. The black line is the data smoothed with a 3 pixel boxcar. The green, red, blue line represents the computed model spectrum with [C]/[He] = -6.4, -6.7, -7.0, respectively, placed at a different continuum level; the abundances of other elements are from Table \ref{Tab: AbundanceGD}. The adopted carbon abundance is -6.7 $\pm$ 0.3. The red labels represent lines that are used for abundance determinations. Wavelength is presented in the star's reference frame in vacuum. } \label{Fig: GD_C}
\plotone{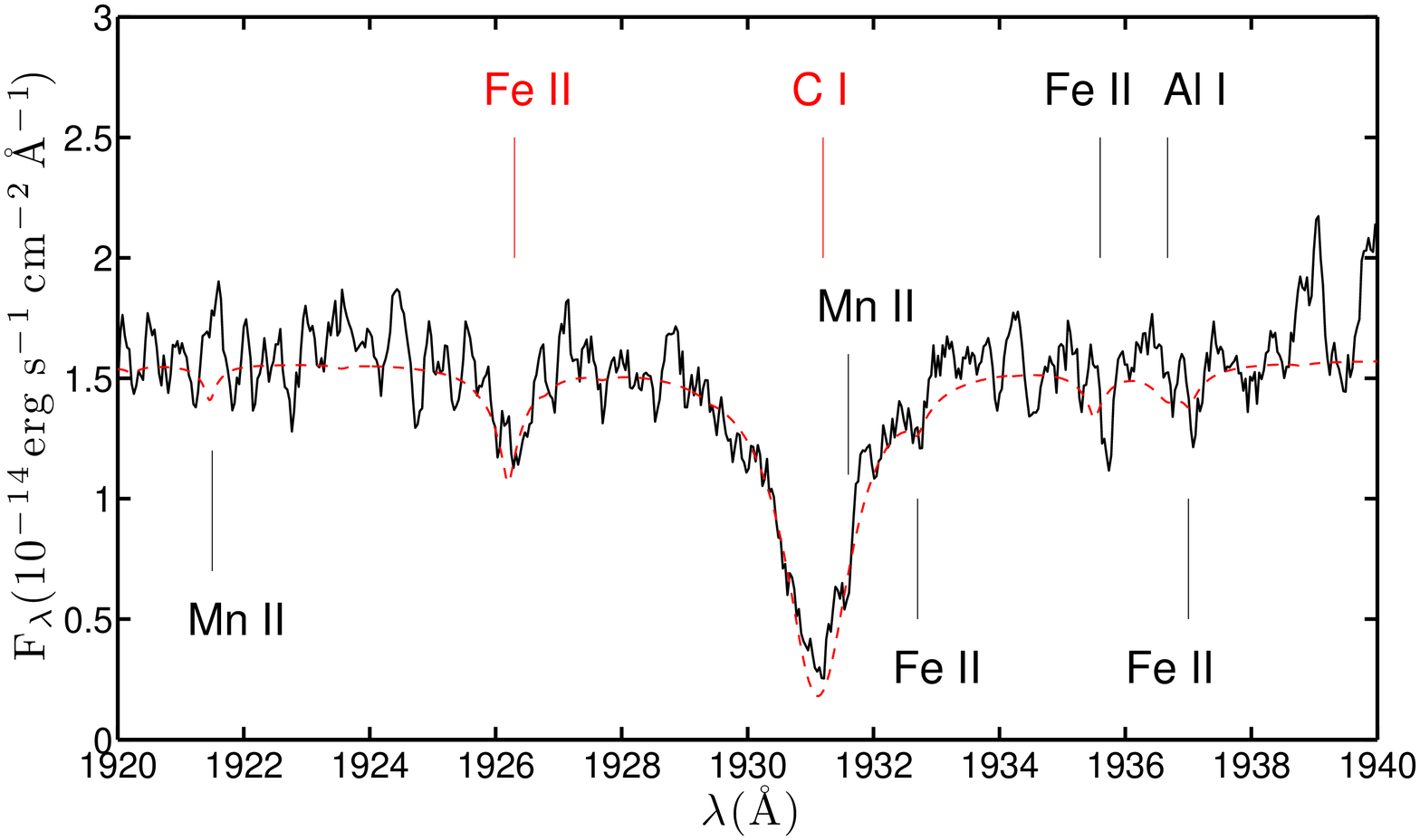}
\caption{Similar to Figure \ref{Fig: GD_C} except for PG 1225-079 with abundances from Table \ref{Tab: AbundancePG}. The data were smoothed with a 5 pixel boxcar.\label{Fig: PG_C}}
\end{figure}

There is only one useful carbon line in the observed wavelength interval, C I 1930.9 {\AA}, as shown in Figures \ref{Fig: GD_C} and \ref{Fig: PG_C}. Because it arises from an excited level, it cannot be contaminated by interstellar absorption. However, this line can be blended with Mn II 1931.4 {\AA}. Fortunately, accurate Mn abundances have been determined for both stars from optical data \citep{Zuckerman2007, Klein2011} and the predicted EW for Mn II 1931.4 {\AA} is less than 50 m{\AA} in the model spectrum. Considering the measured EW of this feature is more than 500 m{\AA} for both stars (see Tables \ref{Tab: LinesGD} and \ref{Tab: LinesPG}), we conclude that the line is dominated by C I 1930.9 {\AA}. For PG 1225-079, our derived carbon abundance\footnote{Here, log n(X)/n(Y) is abbreviated as [X]/[Y].} [C]/[He] = -7.80 $\pm$ 0.10 agrees with the IUE upper limit of -7.5 \citep{Wolff2002}. 

For GD 362, the largest uncertainty is from the low SNR of the data; the measured continuum flux is (3.1 $\pm$ 1.0) $\times$ 10$^{-15}$ erg s$^{-1}$ cm$^{-2}$ {\AA}$^{-1}$. It is hard to measure the EW of C I 1930.9 {\AA} directly from the noisy data. Instead, we computed model spectra with different carbon abundance to match the observed spectrum. In Figure \ref{Fig: GD_C}, we present three best-fit models with [C]/[He] = - 6.4, [C]/[He] = -6.7, [C]/[He] = -7.0 and a continuum flux at 4.1 $\times$ 10$^{-15}$ erg s$^{-1}$ cm$^{-2}$ {\AA}$^{-1}$, 3.1 $\times$ 10$^{-15}$ erg s$^{-1}$ cm$^{-2}$ {\AA}$^{-1}$, 2.1 $\times$ 10$^{-15}$ erg s$^{-1}$ cm$^{-2}$ {\AA}$^{-1}$, respectively. The final abundance is [C]/[He] = -6.7 $\pm$ 0.3 and the EW reported in Table \ref{Tab: LinesGD} is measured from the model spectra.

\subsection{Sulfur}

\begin{figure}[hp]
\epsscale{1.0}
\plotone{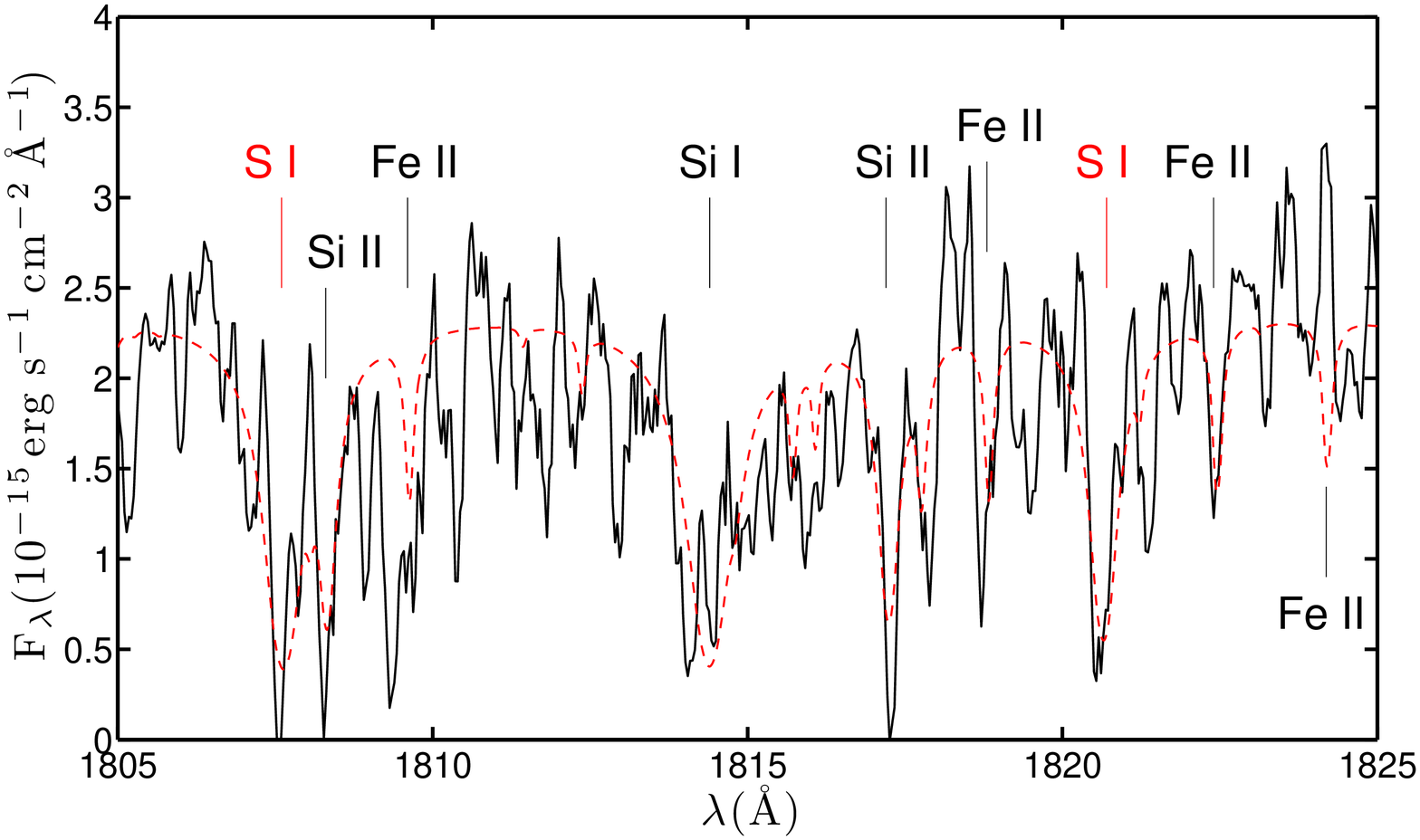}
\caption{{\it HST}/COS spectrum of GD 362. All notations are the same as Figure \ref{Fig: GD_C} and the data are smoothed by a 5 pixel boxcar. S I 1807.3 {\AA} and 1820.3 {\AA} are used for constraining the sulfur abundance.} \label{Fig: GD_S}
\plotone{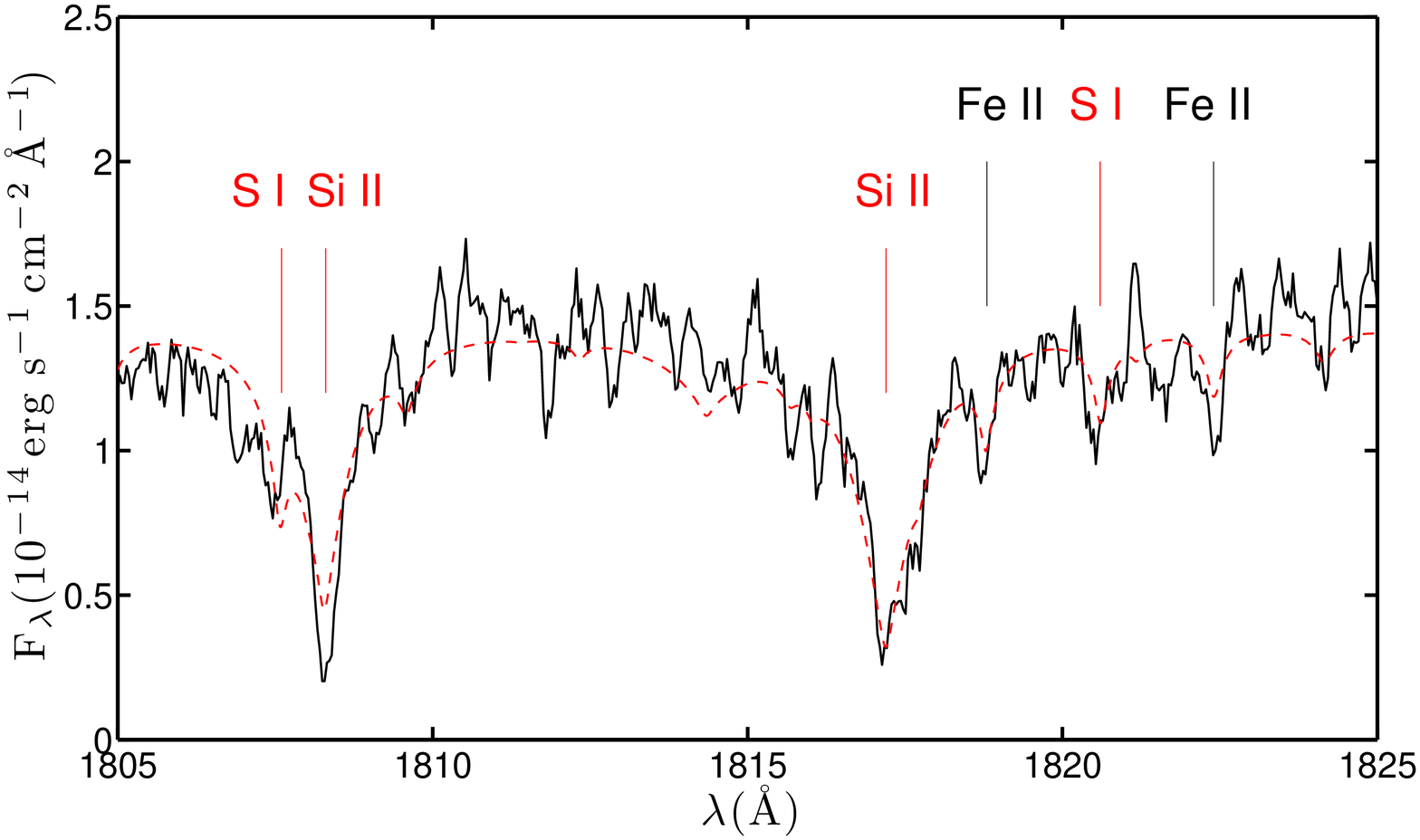}
\caption{Similar to Figure \ref{Fig: GD_S} except for PG 1225-079. Si II 1808.0 {\AA} and 1816.9 {\AA} lines are used for determining the silicon abundance. \label{Fig: PG_S}}
\end{figure}

There are two useful sulfur lines, S I 1807.3 {\AA} and S I 1820.3 {\AA}. However, at best, we have only a tentative detection of sulfur in each star. S I 1807.3 {\AA}, the stronger line, is adjacent to Si II 1808.0 {\AA}. Fortunately, for GD 362, the silicon abundance is determined from previous optical data \citep{Zuckerman2007}; for PG 1225-079, other ultraviolet lines can be used to derive the silicon abundance (see section 3.4). The data and model atmosphere spectra for GD 362 and PG 1225-079 are presented in Figures \ref{Fig: GD_S} and \ref{Fig: PG_S}, respectively. Considering the apparent match between the model and data for both S I lines, tentative sulfur abundances of -6.7 for GD 362 and -9.5 for PG 1225-079 can be assigned. Conservatively, these results are upper limits.

\subsection{Magnesium and Zinc}

\begin{figure}[ht!]
\plotone{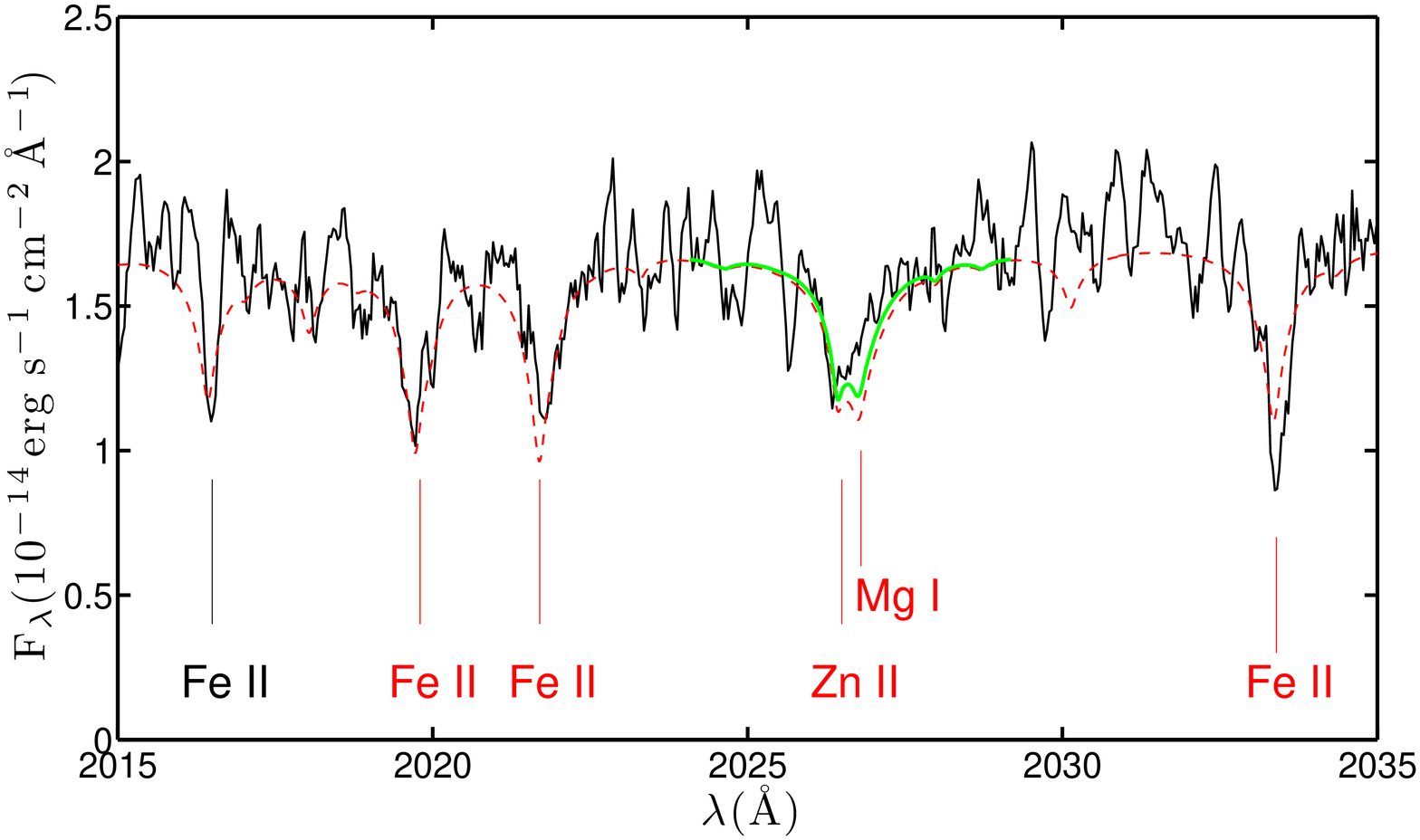}
\caption{{\it HST}/COS spectrum of PG 1225-079 and the data are smoothed by a 5 pixel boxcar. All notations are the same as Figure \ref{Fig: GD_C}. Zn II 2026.1 {\AA} and Mg I 2026.5 {\AA} are used for determining the Zn and Mg abundances, respectively. The green line presents the best fit model that matches the overall EW of the absorption feature at 2026 {\AA} ([Mg]/[He] = -7.6, [Zn]/[He] = -11.3); the red line is the adopted model combining with the optical results ([Mg]/[He] = -7.5, [Zn]/[He] = -11.3). Fe II 2019.4 {\AA}, 2021.4 {\AA}, 2033.0 {\AA} and 1926.0 {\AA} in Figure \ref{Fig: PG_C} are used for determining the iron abundance. \label{Fig: PG_Zn}}
\end{figure}

In PG 1225-079, Mg I 2026.4 {\AA} and Zn II 2026.1 {\AA} are heavily blended. As shown in Figure \ref{Fig: PG_Zn}, our best fit model which matches the measured EW of the absorption feature requires [Mg]/[He] = -7.6 and [Zn]/[He] = -11.3. These values are individually taken as upper limits due to the blending. However, the reported magnesium abundance is -7.27 $\pm$ 0.06 from the optical data \citep{Klein2011}, which is largely based on three Mg lines but the detections for two lines are only 2$\sigma$. \citet{Wolff2002} reported [Mg]/[He] to be -7.6 $\pm$ 0.6 from the IUE data. Averaging these measurements, our final magnesium abundance is -7.50 $\pm$ 0.20. Because of the blending, the zinc abundance is only an upper limit. This provides the first stringent constraint on zinc in an extrasolar planetesimal.

\subsection{Silicon}

In PG 1225-079, we measured two silicon lines, Si II 1808.0 {\AA} and Si II 1816.9 {\AA}, as shown in Figure \ref{Fig: PG_S}. Si II 1808.8 {\AA} arises from the ground state and the photospheric line can be distorted by interstellar absorption. However, its measured EW is only 87 $\pm$ 11 m{\AA} in $\zeta$ Oph, a star at a distance of 112 pc with a large amount of foreground interstellar gas \citep{Morton1975}. Considering PG 1225-079 is only 26 pc away, it has much less interstellar absorption. The measured EW is 936 $\pm$ 109 m{\AA} and we conclude that Si II 1808.0 {\AA} is largely photospheric and essentially free from interstellar absorption. The shape of Si II 1808.0 {\AA} in the model does not quite fit the data; but the measured EW of the data, which is key in the abundance determination, has a good agreement with that in the model. Using these two Si II lines, we derive a final silicon abundance of -7.45 $\pm$ 0.10, in agreement with, but much better than the reported IUE abundance of -7.5 $\pm$ 0.5 \citep{Wolff2002} and the previous optical upper limit of -7.27 \citep{Klein2011}.

\subsection{Iron}

In the COS data for PG 1225-079, there are six Fe II lines with EWs larger than 100 m{\AA}. Four of them are shown in Figures \ref{Fig: PG_C} and \ref{Fig: PG_Zn}. We derived an iron abundance of -7.45 $\pm$ 0.23, in good agreement of the optical value of -7.42 $\pm$ 0.07, which is based on 28 high-SNR iron lines \citep{Klein2011}. Because the ultraviolet data are noisier, we adopt the optically-derived iron abundance.

\section{COMPARISON WITH SOLAR SYSTEM OBJECTS}

Combined with previous data, we now have determined the abundances of 16 elements heavier than helium in the atmosphere of GD 362 and 11 heavy elements in PG 1225-079. However, the measured composition need not be identical to the composition of the accreted planetesimal because different elements gravitationally settle at different rates in a white dwarf atmosphere. Three major phases are proposed for a single accretion event: build-up, steady-state and decay \citep{Dupuis1993a, Koester2009a}. 

Because an infrared excess is found for GD 362 and PG 1225-079 \citep{Becklin2005, Kilic2005, Farihi2010b}, the accretion should be either in the build-up or steady-state phase. The timescale for build-up stage is comparable to the settling times \citep{Koester2009a}; it is $\sim$ 10$^5$ yr, for GD 362 and PG 1225-079 (see Tables \ref{Tab: AbundanceGD} and \ref{Tab: AbundancePG}). The rest of the disk-host stage should all be under the steady-state approximation. The dust disk lifetime has been under intensive studies for a few years but the values are still very uncertain, including 10$^5$ yr \citep{Farihi2009, Rafikov2011b}, 10$^6$ yr \citep{Rafikov2011a, Girven2012, Farihi2012b} and up to 10$^7$ yr \citep{Barber2012}. The true disk lifetime might have a range but it is likely to be longer than the settling times. Furthermore, \citet{Zuckerman2010} suggested that steady-state approximation is the dominant situation for white dwarf accretion event based on a study of helium dominated stars; the settling times are only 0.1\% of their cooling times but 30\% of them show atmospheric pollution. GD 362 and PG 1225-079 are more likely to be under the steady-state approximation and that is the main focus of this paper.

In the steady-state model, the observed concentration of an element is dependent on the time it takes to sink out of the convective envelope. To derive the theoretical settling times and obtain an improved understanding of the uncertainties, we formulated several numerical experiments with the code for the envelope structure and corrected two errors found in our previous calculations of diffusion timescales. In the course of changing the equations describing element diffusion from the version in \citet{Paquette1986} (Equation 4) to the one in \citet{Pelletier1986} (Equation 5), which is more accurate in the case of electron degeneracy, one of us (D.K.) discovered an error in the former paper. A factor of $\rho^{1/3}$ is missing in the second alternative of Equation 21, which we had not noticed before. A rederivation of all our equations uncovered another error in our implementation of the contribution of thermal diffusion. These errors have only a very small effect in stars with relatively shallow convection zones, like the hydrogen-dominated white dwarfs. However, for helium-dominated white dwarfs with T $<$ 15,000 K and a deep convection zone, the diffusion timescales can be slower by factors 2-3 relative to our earlier calculations\footnote{Updated diffusion timescales can be obtained at http://www.astrophysik.uni-kiel.de/~koester/astrophysics/ }. The accretion rate $\dot{M}(Z_i)$ of an element Z is calculated as \citep{Koester2009a}

 \begin{equation}
 \dot{M}(Z_i) = \frac{M_{cvz} X(Z_i)}{t_{set}(Z_i)}
 \end{equation}
where M$_{cvz}$ is the mass of the convective envelope. X(Z$_i$) is the mass fraction of the element Z$_i$ relative to the dominant element in the atmosphere, either hydrogen or helium; t$_{set}$(Z$_i$) is the settling time. A longer settling time corresponds to a lower diffusion flux. Fortunately, the relative timescales for different elements, which are important for the determination of the abundances in the accreted matter, change much less.

For GD 362 and PG 1225-079, compared to previously published values, the settling times listed in Tables \ref{Tab: AbundanceGD} and \ref{Tab: AbundancePG} typically increase by factors of 2-3 while the mass of the convective zone is 0.13 dex smaller for GD 362 and 0.05 dex larger for PG 1225-079 (Table \ref{Tab: Properties}). These corrections lead to smaller total accretion rates by a factor of 3 for both stars. 

The next step is to compare the composition of the accreted parent body with those of solar system objects. We choose the summed mass of all the major elements as the normalization factor so that the analysis is independent of the chemical property and abundance uncertainty of each individual element. However, one complication is that no oxygen lines are detected in either GD 362 or PG 1225-079 due to their low photospheric temperatures relative to other helium-dominated white dwarfs; only upper limits were obtained for this major element. Therefore, our approach is to compare the mass fraction of an element relative to the summed mass of the common elements Mg, Si and Fe. For solar system objects, we include 80 representative and well-analyzed meteorite samples mostly from \citet{Nittler2004}. We also include the bulk composition of Earth from \citet{Allegre2001} and an updated carbon abundance from \citet{Marty2012}. For our purpose, Earth appears to be chondritic and its bulk composition approaches CV chondrites even though Earth has experienced some post-nebular processing, such as differentiation and collisions.

\subsection{GD 362: Accretion from a Mesosiderite Analog?}

\begin{figure}[ht!]
\epsscale{1.0}
\plotone{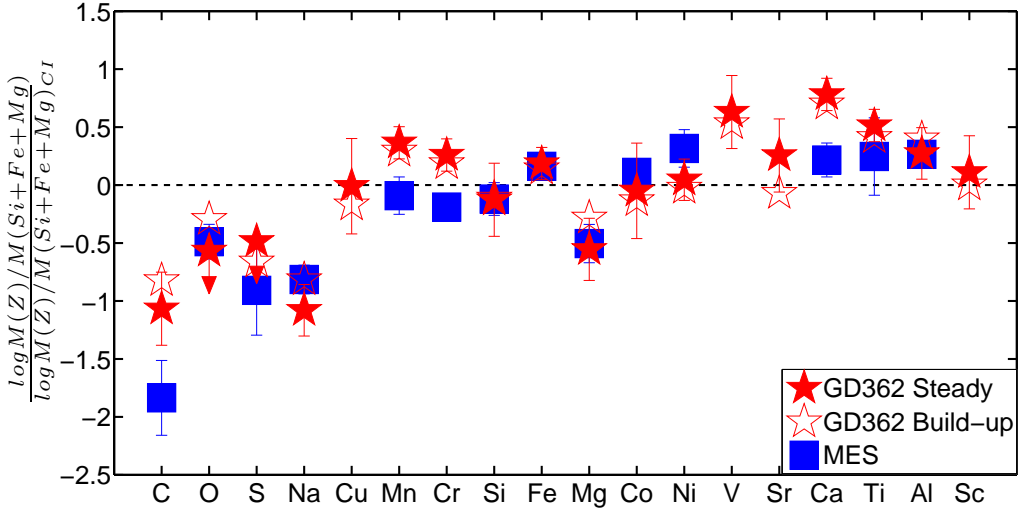}
\caption{Mass fractions of heavy elements in GD 362 with respect to the summed mass of silicon, iron and magnesium from Table \ref{Tab: AbundanceGD}. The abundances are normalized to those of CI chondrites. The elements are ordered by decreasing volatility. The filled stars represent the steady-state approximation while the open stars are in the build-up stage. 1$\sigma$ error bars and arrows for upper limits are plotted only for the steady-state approximation for clarity. We also plot the average values for five mesosiderites (Emergy, Barea, Patwar, Dyarrl Island and ALH 77219) from \citet{Nittler2004}; 1$\sigma$ deviations are shown as blue error bars. The error bars are smaller than the symbol when they are not visible in the figure. There are no reported whole rock fractions of Cu, V, Sr and Sc in mesosiderites.}
\label{Fig: GD362}
\end{figure}

In Figure \ref{Fig: GD362}, we compare the abundances of all 18 elements, including upper limits, of the accreted material in GD 362 with CI chondrites, which are the most primitive material in the solar system. The composition of CI chondrites is almost identical to the solar photosphere, with the exception of depletion of volatile elements C, N as well as H and noble gases. The parent body accreted onto GD 362 looks nothing like a CI chondrite, as first pointed out in \citet{Zuckerman2007}. For the volatile elements, the mass fraction of C and S are depleted by at least a factor of 7 and 3, respectively, relative to CI chondrites; refractory elements, such as V, Ca, Ti and Al, are all enhanced.

Though oxygen is not detected in GD 362, its stringent upper limit can still provide useful insights. Following \citet{Klein2010}, we can calculate the required number of oxygen atoms to form oxides Z$_{p(Z)}$O$_{q(Z)}$ as

\begin{equation}
n(O)=\sum_Z \frac{q(Z)}{p(Z)} n(Z)
\end{equation}
Hydrogen is excluded here because GD 362 has an enormous amount and it might not be associated with the parent body or bodies currently in its atmosphere (see Appendix A). Under the steady-state approximation, [O]/[He] = -5.07 is required to form MgO, Al$_2$O$_3$, SiO$_2$ and CaO; this value is comparable to the observed oxygen upper limit of -5.14. However, Fe is the most abundant heavy element in the atmosphere of GD 362 and there is insufficient oxygen to tie it up in either FeO or Fe$_2$O$_3$. Thus, most, if not all the iron in the parent body is in metallic form, which is very different from CI chondrites where most iron is in oxides \citep{Nittler2004}.

\citet{ONeillPlame2008} suggested that [Mn]/[Na] can be used as an indicator of post-nebular processing. For example, [Mn]/[Na] is -0.79 for all chondrites as well as the solar photosphere while non-chondritic objects have a much higher value. Interestingly, [Mn]/[Na] is 0.65 $\pm$ 0.22 for GD 362, which is larger than -0.01 for Mars and 0.32 for the Moon \citep{ONeillPlame2008}. This suggests that the planetesimal accreted onto GD 362 is likely to be non-chondritic and have experienced some post-nebular processing. \citet{Zuckerman2007} compared the [Na]/[Ca] ratio in GD 362 with solar system objects and reached a similar conclusion; the accreted planetesimal was non-chonridtic. The only other polluted white dwarf with both Mn and Na detections is WD J0738+1835 wherein [Mn]/[Na]= -0.54 $\pm$ 0.19 \citep{Dufour2012}; this agrees with the chondritic value within the uncertainties.

To find the best solar system analog to the parent body accreted onto GD 362, we calculated a reduced chi-squared value for each object in our sample ($\chi^2_{red}$), defined as:

\begin{equation}
\chi^2_{red}= \frac{1}{N} \sum ^N _{i=1} \frac{(M_{wd}(Z_i)-M_{mtr}(Z_i))^2}{\sigma_{wd}^2(Z_i)}
\end{equation}
where N is the total number of elements considered in the analysis. M$_{wd}$(Z$_i$) and M$_{mtr}$(Z$_i$) represent the mass fraction of an element Z$_i$ relative to the summed mass of Mg, Si and Fe in the extrasolar planetesimal and solar system objects, respectively. $\sigma_{wd}$(Z$_i$) is the propagated uncertainty in mass fraction. 

\begin{figure}[hp]
\epsscale{1.1}
\plotone{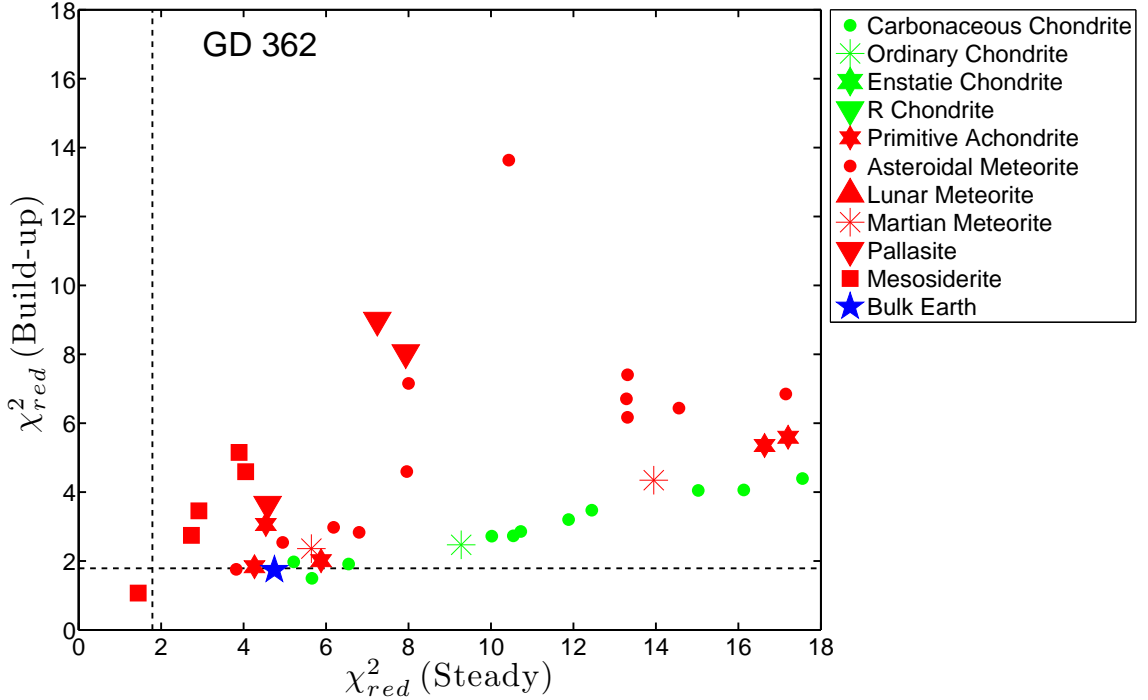}
\caption{This figure shows $\chi_{red}^2$ values defined in Equation (3), which compares the abundances of C, Na, Mg, Al, Si, Ca, Ti, Cr, Mn, Fe and Ni in the accreted material in GD 362  with 80 meteorites and bulk Earth. The x-axis denotes $\chi^2_{red}$ calculated for the composition under the steady-state approximation while the y-axis is for the build-up phase. The dashed lines represent 95\% confidence level. There are a couple of meteorites with a large $\chi^2_{red}$ and they are not shown in the current scale. There is no qualitative difference between the steady-state and build-up models. Mesosiderites, particuarly ALH 77219, provide the best fit to the abundance pattern observed in GD 362. {\bf Notes}: ``Chondritic" materials include: (i) carbonaceous chondrites: CI, CK, CM, CO, CR and CV; (ii) ordinary chondrites: H, L and LL; (iii) enstatite chondrites: EH and EL; (iv) R chondrite. ``Non-chondritic" materials consist of achondrites and stoney-iron meteorites. No iron meteorite is included in our analysis because their bulk composition is dominated by Fe and Ni, with very few trace elements reported. Achondrites include: (i) primitive achondrites: acapulcoites, lodranites, winonaites and ureilites; (ii) asteroidal meteorites: angrites, aubrites, brachinites and howardite-eucrite-diogenite; (iii) Martian meteorites: shergottie, Nakhlites and Chassignites; (iv) Lunar meteorites. For stoney-iron meteorites, we include (i) mesosiderites; (ii) pallasites. Most of the meteorite data are from \citet{Nittler2004} and the compositions for some Martian meteorites are from \citet{McSween1985}. The bulk composition of Earth is from \citet{Allegre2001} and the carbon abundance is from a more recent study of \citet{Marty2012}, which is a factor of 3 lower than the lower limit reported in \citet{Allegre2001}.}
\label{Fig: GD_chi}
\end{figure}

For GD 362, we calculated $\chi^2_{red}$ for 11 heavy elements, C, Na, Mg, Al, Si, Ca, Ti, Cr, Mn, Fe and Ni, which have detections both in GD 362 and the meteorite sample\footnote{For a couple of meteorites with no reported carbon abundance, we compute the $\chi^2_{red}$ for the other 10 elements.}. The results are shown in Figure \ref{Fig: GD_chi} for both steady-state and build-up approximations. There is no qualitative difference between these two models and mesosiderites provide the best fit considering all 11 elements. In particular, the mesosiderite ALH 77219 can match the overall abundance pattern to 95\% confidence level. As shown in Figure \ref{Fig: GD362}, the abundance of individual elements agrees within 2$\sigma$ between mesosiderites and the planetesimal accreted onto GD 362. 

Mesosiderites are a rare type of stoney-iron meteorite with equal amounts of silicates and metallic iron and nickel. One mystery about mesosiderties is that the Si-rich crust and Fe, Ni-rich core materials are abundant but the olivine Mg-rich mantle seems to be missing.  One model for the formation of mesosiderites is that a 200-400 km diameter asteroid with a molten core was nearly catastrophically disrupted by a 50-150 km diameter projectile at 4.42-4.52 Gyr ago \citep{Scott2001}. The collision mixed the target's molten core with its crustal material but excluded the large and hot mantle fragments. The planetesimal accreted onto GD 362 may have been formed in a similar way. 

While mesosiderites may be a prototype for the accreted planetesimal onto GD 362, there are three major hurdles for this hypothesis to overcome. First, in the model of \citet{Scott2001}, only half of the original mass of a 200-400 km diameter asteroid was maintained after the collision and the final product only contains about 10\% mesosiderite-like material by mass. This is equivalent to a 75-150 km diameter object. Mesosiderites that fall on Earth are only small fragments and the 180 kg NWA 2924 is among the largest (Meteorite Bulletin Database\footnote{http://www.lpi.usra.edu/meteor/}). However, the parent body accreted onto GD 362 has a minimum mass of 2.7 $\times$ 10$^{22}$ g, $\sim$260 km in diameter for an assumed density of 3 g cm$^{-3}$. It is unclear whether the same kind of collision can produce a mesosiderite parent body this big. Second, the mass fraction of hydrogen in mesosiderites is less than 0.2\%; it cannot explain how there is 5 $\times$ 10$^{24}$ g hydrogen in the atmosphere of GD 362. Possibly, hydrogen was accreted during earlier events and it has been atop the atmosphere ever since (see Appendix A for more discussion). Third, GD 362 is currently accreting from its circumstellar disk and the disk material should also resemble the composition of mesosiderites. However, the shape of the mid-infrared spectrum for mesosiderite, which is dominated by a sharp peak at 9.13 $\mu$m and several other bands at 10.6 $\mu$m and 11.3 $\mu$m \citep{Morlok2012}, cannot fully account for the broad 10 $\mu$m silicate emission feature observed for GD 362 \citep{Jura2007a}. This does not completely exclude the mesosiderite hypothesis but emission from some additional material is required to fully reproduce the observed infrared spectrum for GD 362. Mesosiderites are a good candidate for the parent body accreted onto GD 362 but there are remaining unresolved issues.

\subsection{PG 1225-079: Accretion from a Planetesimal with No Single Solar System Analog}

\begin{figure}[hp]
\epsscale{1.0}
\plotone{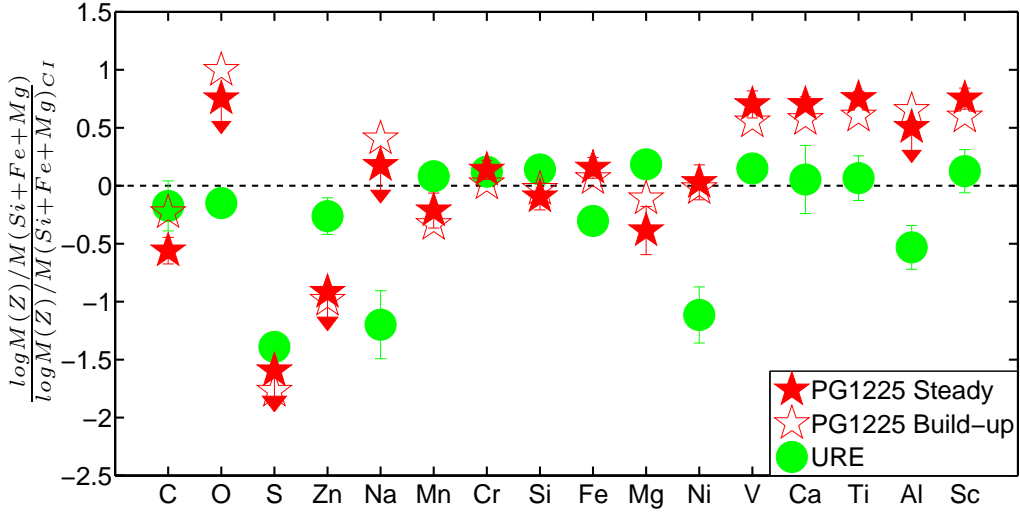}
(a)
\plotone{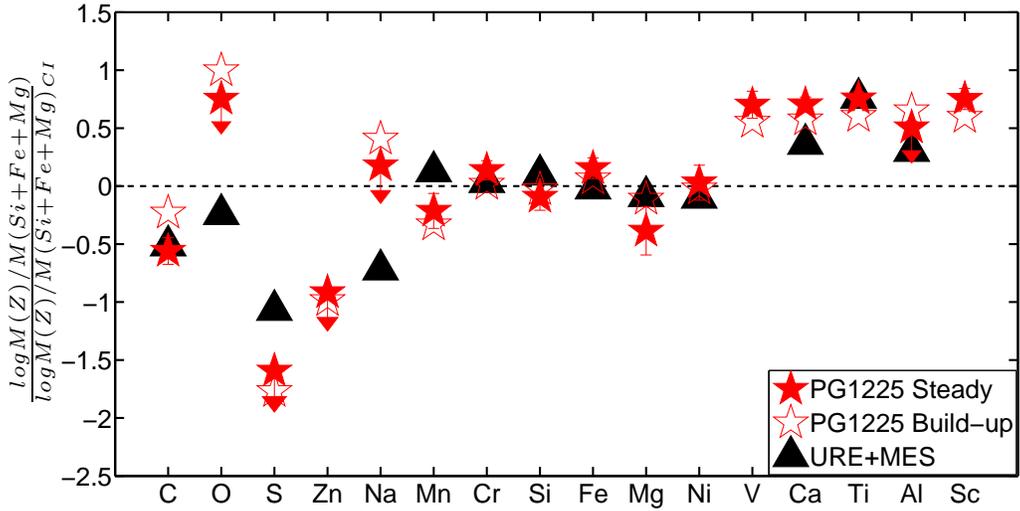}
(b)
\caption{(a) Similar to Figure \ref{Fig: GD362} except for PG 1225-079 with the abundances in Table \ref{Tab: AbundancePG}. Ureilites can match the high carbon low sulfur pattern in PG 1225-079 but fail with the other elements. The abundances for ureilites are from \citet{Warren2006}. (b) The best-fit model for the composition in PG 1225-069 under the steady-state approximation is 30\% by mass ureilite North Haig and 70\% mesosiderite Dyarrl Island. There are no reported bulk compositions of Zn, V or Sc for the North Haig and Dyarrl Island meteorites.
}
\label{Fig: PG1225}
\end{figure}

In Figure \ref{Fig: PG1225}, we show a comparison of  the mass fractions of 16 elements, including upper limits between PG 1225-079 and CI chondrites. Though the carbon abundance is approaching the chondritic value, the accreted planetesimal differs a lot from CI chondrites; the mass fraction of S is depleted by at least a factor of 40 while Zn is depleted by at least a factor of 8. In contrast, refractories, such as V, Ca, Ti and Sc are all enhanced. The overall pattern of relatively high carbon abundance and enhanced mass fractions of refractory elements does not follow a single condensation sequence and post-nebular processing is required.

\begin{figure}[hp]
\plotone{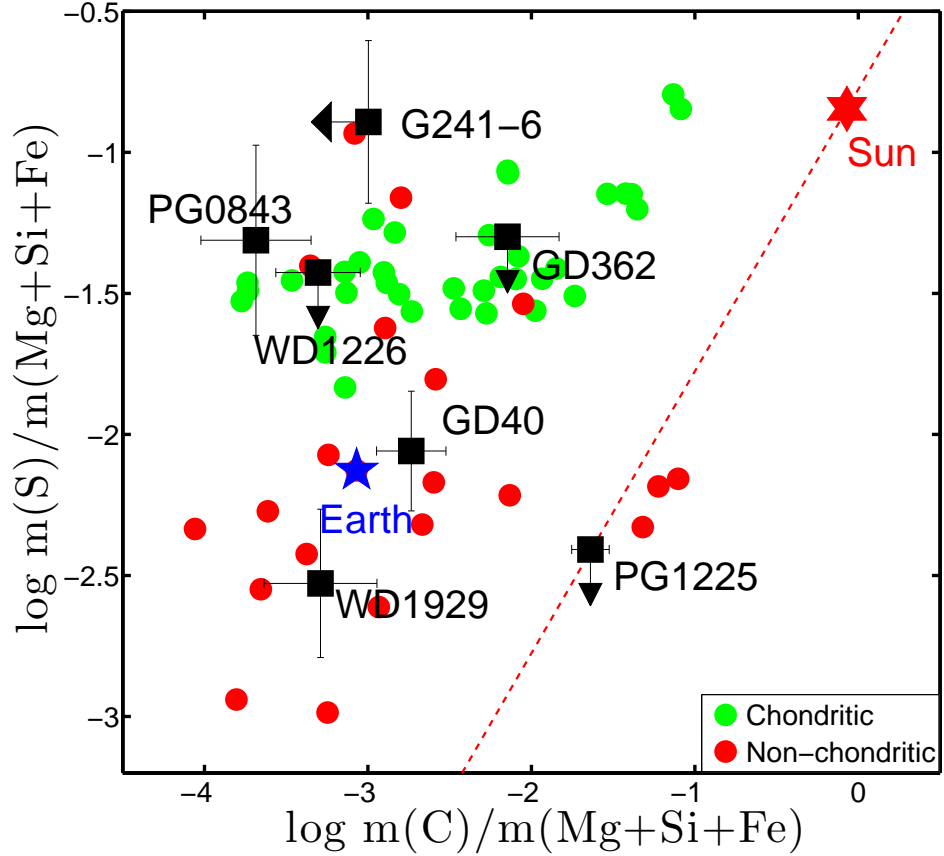}
\caption{Mass fraction of S and C over the sum of Fe, Mg and Si for solar system objects as well as polluted white dwarfs with positive detection of carbon or sulfur or both. Due to the presence of an infrared excess except for G241-6 (see discussion in Appendix B), all white dwarfs are plotted under the steady-state approximation . 1$\sigma$ uncertainties are plotted. The red dashed line denotes constant solar [C]/[S]. Most meteorites have a lower [C]/[S] than the solar value. However, [C]/[S] in PG 1225-079 is no smaller than solar and the closest solar system analog is ureilites. {\bf References}: WD 1929+012, PG 0843+517, WD 1226+110: \citet{Gaensicke2012}; GD 40, G241-6: \citet{Jura2012}; GD 362 and PG 1225-079: this paper; Solar abundance: \citet{Lodders2003}; for solar system objects, the references are listed in the caption of Figure \ref{Fig: GD_chi}.}
\label{Fig: C-Si-S}
\end{figure}

As shown in Figure \ref{Fig: C-Si-S}, PG 1225-079 has a [C]/[S] value that is no smaller than the solar ratio, which is very different from other polluted white dwarfs and meteorites. Carbon and sulfur are among the most volatile elements that we can measure and their 50\% condensation temperatures are 40 K and 655 K, respectively \citep{Lodders2003}. Most of the meteorites as well as polluted white dwarfs have a [C]/[S] ratio lower than the solar value, which can be explained by condensation at a temperature between 40 and 665 K though this is not necessarily true for all of them. The only solar system analog to PG 1225-079 with similar high carbon, low sulfur pattern is ureilites, a type of primitive achondrites. Ureilites are the second largest achondrite group and it is suggested that its high carbon abundance is derived from a carbon-rich parent body, but the exact formation mechanism is not well understood \citep{Goodrich1992}. However, as can be seen in Figure \ref{Fig: PG1225}(a), ureilites fail to match the overall composition of the parent body accreted onto PG 1225-079.

\begin{figure}[ht!]
\epsscale{1.1}
\plotone{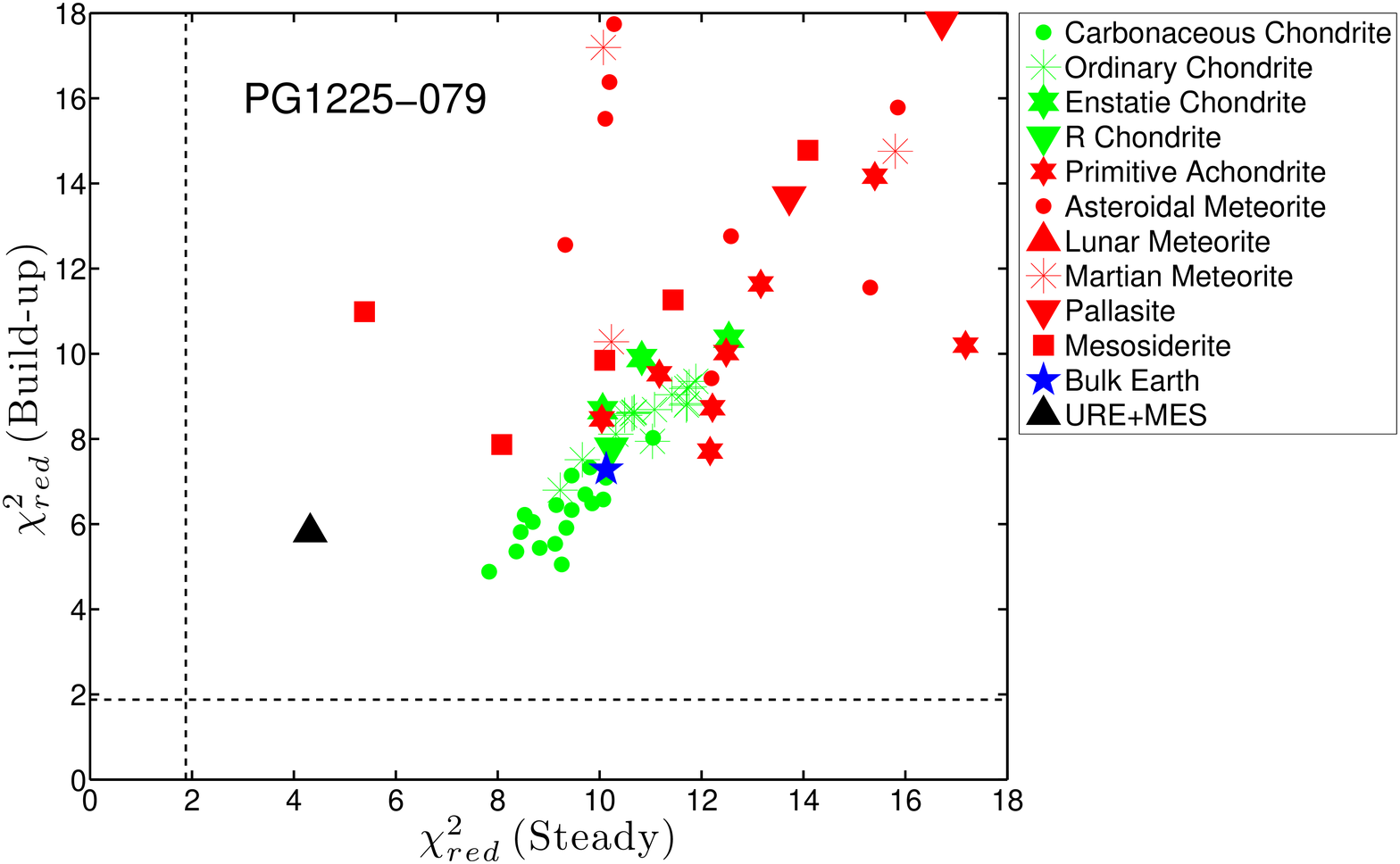}
\caption{Similar to Figure \ref{Fig: GD_chi} except for PG 1225-079 comparing 9 elements, C, Mg, Si, Ca, Ti, Cr, Mn, Fe and Ni. We see that no solar system object comes close to the overall abundance pattern in PG 1225-079. The black triangle represents 30\% by mass ureilite and 70\% mesosiderite as shown in Figure \ref{Fig: PG1225}(b), which is the best match for the composition of PG 1225-079 in the steady-state approximation.}
\label{Fig: PG_chi}
\end{figure}

We performed a $\chi^2_{red}$ analysis between solar system objects and the accreted planetesimal in PG 1225-079, comparing 9 elements, C, Mg, Si, Ca, Ti, Cr, Mn, Fe and Ni\footnote{Similar to the case of GD 362, for the meteorites with no reported carbon abundance, we only calculated $\chi^2_{red}$ for the other 8 elements.}. The result is shown in Figure \ref{Fig: PG_chi}. There is no single solar system object that can match all nine elements; the closest is carbonaceous chondrite. Regardless, as shown in Figure \ref{Fig: PG1225}, the accreted abundance in PG 1225-079 is not at all identical to CI chondrites.

The infrared excess around PG 1225-079 corresponds to $\sim$500 K dust \citep{Farihi2010b}; so far, only two white dwarfs are known to have such cool dust. The other 28 known disk-host stars all have $\sim$1000 K dust \citep{XuJura2012}. One hypothesis is that the inner disk region was recently impacted by another asteroid and all the material was dissipated \citep{Farihi2010b, Jura2008}. If that is the case, PG 1225-079 can be accreting from a blend of two planetesimals, rather than one single parent body. After testing different combinations of the 80 meteorites in our database, the best fit model to the steady-state approximation consists of 30\% ureilite North Haig and 70\% mesosiderite Dyarrl Island by mass. This blend is also marked in Figure \ref{Fig: PG_chi}. Detailed abundance comparison is shown in Figure \ref{Fig: PG1225}(b); the abundances of S, Mn and Ca do not agree as well as the other elements but are all within 2$\sigma$. A possible scenario is that one extrasolar ureilite (mesosiderite) analog first got tidally disrupted and more recently, another mesosiderite (ureilite) analog impacted the disk and was blended with the previous material.

\section{ASSESSING THE FORMATION MECHANISMS OF EXTRASOLAR PLANETESIMALS}

Having established that the parent bodies accreted onto GD 362 and PG 1225-079 are beyond primitive, we now extend our analysis to other extrasolar planetesimals. We are most interested in understanding the formation mechanisms of extrasolar planetesimals and whether these are dominated by nebular or post-nebular processing. \citet{JuraXu2013} suggested collisional rearrangement is important in determining the final composition of extrasolar planetesimals based on the scatter in [Mg]/[Ca] ratios in 60 externally-polluted white dwarfs. Here, we compile a sample of well-studied externally-polluted white dwarfs with abundance determinations of at least 9 elements. There are 9 stars in total, as listed in Table \ref{Tab: WDs} and now we assess the formation mechanism for individual objects.

\begin{sidewaystable}
\begin{center}
\caption{Summary of Planetesimal Formation Mechanisms in 9 Well-Studied White Dwarfs }
\begin{tabular}{lccllllll}
\\
\hline \hline
star	& Dom.	& Dust	& Volatile	& Intermediate & Refractory	& Process	& Ref \\
\hline
GD 40		& He	& Y	& C,S:,O,Mn, P  	& Cr,Si,Fe,Mg,Ni	& Ca,Ti,Al	& primitive	& 1,2\\
WD J0738+1835& He& Y	& O,Na,Mn		& Cr,Si,Fe,Mg,Co,Ni & V,Ca,Ti,Al,Sc	&primitive	& 3,4\\
PG 0843+517	& H	& Y	& C,S,O,P			& Cr,Si,Fe,Mg,Ni	& Al	 	& beyond-primitive(?)	&  5\\
PG 1225-079	& He	& Y	& C,Mn        		& Cr,Si,Fe,Mg,Ni  	& V,Ca,Ti,Sc	& beyond-primitive	&  6,7 \\
NLTT 43806	& H	& N	& Na	 			& Cr,Si,Fe,Mg,Ni 	& Ca,Ti,Al	&  beyond-primitive	& 8\\
GD 362		& He	& Y	& C,Na,Cu,Mn 		& Cr,Si,Fe,Mg,Co,Ni	& V,Sr,Ca,Ti,Al,Sc 	& beyond-primitive	& 7,9\\
WD 1929+012	& H	& Y	& C,S,O,Mn,P		& Cr,Si,Fe,Mg,Ni 	& Ca,Al		& ???	& 5,10,11,12\\
G241-6		& He	& N	& S,O,Mn,P  		& Cr,Si,Fe,Mg,Ni	& Ca,Ti		& primitive	& 2,6,13\\
HS 2253+8023	& He	& N	& O,Mn  			& Cr,Si,Fe,Mg,Ni:	& Ca,Ti	& primitive 	& 6\\
\hline
\end{tabular} \label{Tab: WDs}
\end{center}
{\bf Note.} This is a compiled sample of externally polluted white dwarfs with detections of at least 9 elements heavier than helium. Columns are defined as follows. ``Dom" lists the dominant element in the atmosphere. ``Dust" indicates whether a star has an infrared excess (``Y") or not (``N"). Following the classification scheme in \citet{Lodders2003}, ``Volatile" lists the detected volatile elements, defined as having a 50\% condensation temperature lower than 1290 K in a solar-system composition gas \citep{Lodders2003}; ``Intermediate" lists the elements with a condensation temperature between 1290-1360 K --the same range as that of the common elements, Si, Fe and Mg; ``Refractory" elements have a 50\% condensation temperature higher than 1360 K. The elements are ordered with increasing condensation temperature. ``Process" shows our proposed dominant mechanism that determines the final composition of the accreted extrasolar planetesimal (see section 5). \\
{\bf References.} (1) \citet{Klein2010}; (2) \citet{Jura2012}; (3) \citet{Dufour2010}; (4) \citet{Dufour2012}; (5) \citet{Gaensicke2012}; (6) \citet{Klein2011}; (7) this paper; (8) \citet{Zuckerman2011}; (9) \citet{Zuckerman2007}; (10) \citet{Vennes2010}; (11) \citet{Vennes2011a}; (12) \citet{Melis2011}; (13) \citet{Zuckerman2010}.
\nl
\end{sidewaystable}

{\it GD 40}: As discussed in \citet{Jura2012} and Appendix B, the overall abundance pattern in GD 40 matches with carbonaceous chondrites and bulk Earth. Nebular condensation is sufficient to explain its observed composition.

{\it WD J0738+1835}: \citet{Dufour2012} found that there is a correlation between the abundance of an element and its condensation temperature: refractory elements are depleted while volatile elements are enhanced compared to bulk Earth. This indicates that the accreted planetesimal might be formed in a low temperature environment under nebular condensation.

{\it PG 0843+517}: This star has the  highest mass fraction of iron among all polluted white dwarfs. \cite{Gaensicke2012} found that all core elements, including Fe, Ni, S and Cr are enhanced relative to the values for bulk Earth while lithophile refractory Al is depleted. This star might be accreting from the core of a differentiated object. Nevertheless, considering the uncertainty for each element is at least 0.2 dex, the conclusion is still preliminary.

{\it PG 1225-079}: As discussed in section 4.2, this star has a near chondritic carbon abundance but also enhanced mass fractions of refractory elements relative to CI chondrite; it  cannot be formed solely under nebular processing.

{\it NLTT 43806}: Compared to chondritic values, the accreted planetesimal is depleted in Fe and enhanced in Al. \citet{Zuckerman2011} found that the best fit model corresponds to ``30\% crust 70\% upper mantle". With detections of 9 elements, evidence is strong that NLTT 43806 has accreted the outer layer of a differentiated parent body.

{\it GD 362}: As discussed in section 4.1, mesosiderite is the best solar system analog to the accreted parent body and post-nebular processing is required.

{\it WD 1929+012}: \citet{Gaensicke2012} showed that this star has a high iron content. However, the situation is perplexing in that different analyses yield different stellar parameters and atmospheric abundances. For example, both \citet{Melis2011} and \citet{Gaensicke2012} derived that [Si]/[Fe] is -0.25 but \citet{Vennes2010} found that [Si]/[Fe] is 0.19. No final conclusion can be drawn before resolving such discrepancies.

{\it G241-6}: This star is a near twin of GD 40 with a similar abundance pattern but without an infrared excess. One possible scenario is that G241-6 has accreted a planetesimal with a similar composition to GD 40 and now it is at the beginning of a decaying phase; all heavier elements appear to be depleted relative to GD 40 due to their short settling times \citep{Klein2011,Jura2012}. As discussed in \citet{Jura2012} and Appendix B, the overall abundances resemble those of chondrites and no post-nebular processing is required.

{\it HS 2253+8023}: \citet{Klein2011} showed that the composition of its parent body agrees with bulk Earth, except for the enhanced calcium abundance. Nebular processing can produce the observed abundance pattern.

As summarized in Table \ref{Tab: WDs}, at least 4 out of the 9 white dwarfs have accreted planetesimals that can be formed under nebular processing while post-nebular processing is required for another 3 of them. It should be noted that some objects that we identify as primitive might still have undergone some post-nebular processing. For example, GD 40 has accreted from a planetesimal that has a similar composition as bulk Earth, whose overall abundance pattern is chondritic. However, it is still possible that the parent body was differentiated; when the entire object is accreted, the composition appears to be ``chondritic". We can only put an upper limit on the number of objects formed under nebular condensation. 

From this sample of 9 stars, we see that post-nebular processing appears to play an important role in determining the final abundance of extrasolar planetesimals; beyond-primitive planetesimals might be as common as primitive planetesimals. In contrast, chondrites comprise more than 90\% of all meteorites found on Earth by number (Meteorite Bulletin Database\footnote{http://www.lpi.usra.edu/meteor/}). Possibly, extrasolar planetesimals around white dwarfs have violent evolutionary histories with more collisions. This difference is not surprising since dynamical rearrangement of planetary systems at white dwarfs is expected to increase the frequency of collisions and produce more beyond-primitive extrasolar planetesimals.

So far, 19 elements heavier than helium, including C, S, O, Na, Cu, Mn, P, Cr, Si, Mg, Fe, Co, Ni, V, Sr, Ca, Ti, Al and Sc, have been detected in the atmospheres of polluted white dwarfs, as shown in Table \ref{Tab: WDs}. In terms of mass fraction in the accreted planetesimal, the lowest limit is $\sim$5 ppm, for Sc in WD J0738+1835 \citep{Dufour2012}. Studying externally-polluted white dwarfs proves to be a very sensitive probe of the bulk compositions of extrasolar planetesimals.

\section{CONCLUSIONS}

We present {\it HST}/COS ultraviolet observations for GD 362 and PG 1225-079, two heavily polluted  helium white dwarfs. In GD 362, the mass fractions of carbon and sulfur are depleted by at least a factor of 7 and 3 respectively, compared to CI chondrites. In PG 1225-079, a similar volatile depletion pattern is found: C by a factor of 2, S by at least a factor of 40 and Zn by at least a factor of 8. We provide good evidence for the presence of beyond-primitive extrasolar planetesimals:

\begin{enumerate}

\item Mesosiderites provide a good match to the composition of the parent body accreted onto GD 362. However, there are several unresolved issues for this hypothesis, especially the apparent difference between the mid-infrared spectrum of mesosiderites and the dust disk around GD 362. Additional material is required.

\item No single meteorite can reproduce the abundance pattern in PG 1225-079. A blend of 30\% North Haig ureilite and 70\% Dyarrl Island mesosiderite can provide a good fit to the overall composition. 

\item Spectroscopic observations of externally-polluted white dwarfs enable sensitive measurement of the bulk compositions of extrasolar planetesimals, including 19 heavy elements down to a mass fraction of 5 ppm. Based on a sample of 9 well-studied white dwarfs, we find that post-nebular processing is as important as nebular condensation in determining the compositions of extrasolar planetesimals.

\end{enumerate}

Support for program \# 12290 was provided by NASA through a grant from the Space Telescope Science Institute, which is operated by the Association of Universities for Research in Astronomy, Inc., under NASA contract NAS 5-26555. This work also has been partly supported by NSF grants to UCLA to study polluted white dwarfs.

\appendix
\begin{center}
{\bf APPENDIX}
\end{center}
\section{The {\it Herschel}/PACS Observation of GD 362}

While hydrogen is detected in some helium-dominated white dwarfs \citep{Voss2007}, GD 362 has an anomalously large amount. The helium-to-hydrogen number ratio is 14 in its convective zone, corresponding to 5 $\times$ 10$^{24}$ g of hydrogen; this is lower than 7 $\times$ 10$^{24}$ g reported in \citet{Jura2009b} because the mass of the convective zone for GD 362 is 0.13 dex lower in the updated calculation (Table \ref{Tab: Properties}). The origin of the hydrogen is a mystery. Unlike heavy elements which have short settling times compared to the white dwarf cooling age, hydrogen never sinks and can be accumulated over the entire cooling history of the star \citep{Bergeron2011,JuraXu2012}. If GD 362 has always been a helium-dominated white dwarf and all this hydrogen is from accretion of tidally disrupted objects, it can either be one Callisto-size object or $\sim$100 Ceres-like asteroids \citep{Jura2009b}. In the latter case, likely there would be many more asteroids orbiting the star and mutual collisions among them would generate a cloud of cold dust. 

We were awarded 1.1 hours of {\it Herschel}/PACS \citep{Poglitsch2010} observation time to look for cold dust around GD 362. The ``mini-scan map" mode was used to observe in ``blue" (85-125 $\mu$m) and ``red" (125-210 $\mu$m) bands simultaneously with a medium scan speed of 20{\arcsec} s$^{-1}$ and a scan leg length of 4\arcmin. The scan map size is 345{\arcsec} $\times$ 374{\arcsec} and the repetition number is 25. Two different scan angles, 45 degrees and 135 degrees were used and the total integration time was 1200 sec.

Data reduction was performed using HIPE (Herschel Interactive Processing Environment) on a combined mosaic of level 2 products from pipeline SPG 7.1.0. The pixel scale is 1{\arcsec} pixel$^{-1}$ and 2{\arcsec} pixel$^{-1}$ for the blue and red band, respectively. Correcting for its proper motion, we expect GD 362 at $\alpha$ = 17:31:34.355, $\delta$ = +37:05:18.331 on the date of the observation.  Because there is no detection, aperture photometry was performed at 25 locations within 5 pixels of the nominal position of GD 362. The aperture radius was 20{\arcsec} with a sky annulus between 61{\arcsec} and 70{\arcsec}. The background intensity was estimated using the median sky estimation algorithm ({\it Herschel} Data Analysis Guide\footnote{http://herschel.esac.esa.int/hcss-doc-8.0/print/howtos/howtos.pdf}). Aperture correction factors are 0.949 for blue and 0.897 for red (PACS Observer's Manual\footnote{http://herschel.esac.esa.int/Docs/PACS/pdf/pacs\_om.pdf}). Based on the dispersion of the 25 measurements, 3$\sigma$ upper limits are 5.1 mJy for blue and 5.6 mJy for red.

What does this imply about dust mass? GD 362 has shrunk in mass from 3 M$_{\odot}$ on the main-sequence to its current mass of 0.72 M$_{\odot}$ \citep{Kilic2008b}. Consequently, asteroids initially at 3-5 AU are now orbiting at 13-21 AU. Currently, GD 362 has a stellar temperature of 10,540 K and cooling age $\sim$ 0.9 Gyr \citep{Farihi2009}. Extrapolating from white dwarf cooling models\footnote{http://www.astro.umontreal.ca/~bergeron/CoolingModels/} \citep{Bergeron2011}, for GD 362, its stellar temperature is lower than 20,000 K for 90\% of its cooling time. We approximate the stellar luminosity as a time-averaged luminosity of 0.01 L$_\odot$. Poynting-Robertson drag was able to remove particles smaller than 20 $\mu$m at a distance of 15 AU for a grain density of 3 g cm$^{-3}$. We therefore assume a dust particle radius of 20 $\mu$m in the putative asteroid belt orbiting GD 362.

If the grains function as blackbodies with negligible albedo, then their temperature can be calculated as

\begin{equation}
T_d = T_* \sqrt{\frac{R_*}{2D_{orb}}}
\end{equation}

T$_*$, R$_*$ are the stellar temperature and radius; D$_{orb}$ is the orbital distance. The dust temperature is 14-11 K between 13-21 AU.

\begin{figure}[ht!]
\epsscale{0.8}
\plotone{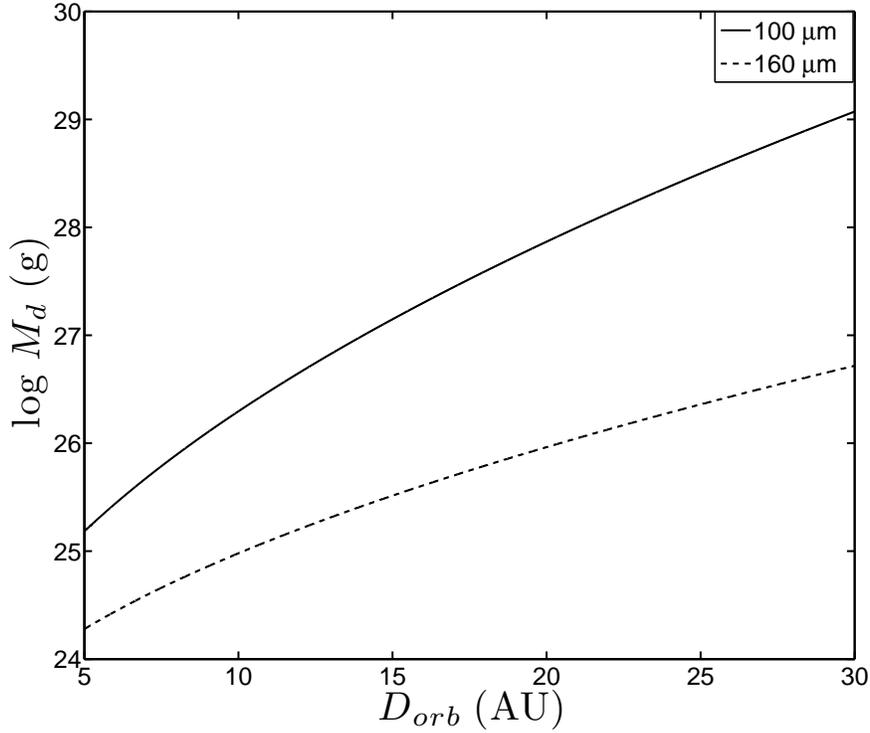}
\caption{Upper limit for the mass of cold dust around GD 362 derived from PACS blue and red data, as a function of orbital radius. Given the hydrogen mass of 5 $\times$ 10$^{24}$ g, the upper limit of dust mass 10$^{25}$-10$^{26}$ g at 13-21 AU is not stringent enough to rule out the accretion of multiple asteroids.}
\label{Fig: MD}
\end{figure}

The mass of the dust disk is

\begin{equation}
M_{d}=\frac{ F_{\nu} D_*^2}{\chi B_{\nu}(T)} 
\end{equation}
where D$_*$ is the distance to GD 362, 51 pc \citep{Kilic2008b} and $\chi$ is the dust opacity. For a particle radius of 20 $\mu$m, $\chi$ = 100 cm$^2$ g$^{-1}$ in the geometric optics limit. As shown in Figure \ref{Fig: MD}, the upper limit of dust mass is between 10$^{25}$ g and 10$^{26}$ g at 13-21 AU; this mass is at least twice the hydrogen mass in the atmosphere of GD 362 and one order of magnitude larger than the mass of solar system's asteroid belt \citep{Krasinsky2002}. The upper limit is not stringent enough to rule out the hypothesis that hydrogen in GD 362 is from accretion of multiple asteroids. So, the large hydrogen abundance in GD 362 remains an unsolved puzzle.

\section{Looking for Solar System Analogs to Extrasolar Planetesimals}

The $\chi^2_{red}$ analysis has proven to be an effective way to look for solar system analogs to the compositions of extrasolar planetesimals. Two other helium-dominated white dwarfs have reported volatile and refractory abundances from high-resolution optical and ultraviolet observations that are suitable for this kind of analysis\footnote{GD 61 also has high-resolution optical and ultraviolet observations \citep{Desharnais2008, Farihi2011a}. However, with a total of 5 detected elements, it is hard to make a comparison using the $\chi^2_{red}$ analysis.} -- GD 40 and G241-6. Updated settling times and accretion rates are listed in Table \ref{Tab: GD40G241-6} while the mass of the convective zone stays the same. Since all the major elements are determined, we compare the mass fraction of an element relative to the sum of O, Mg, Si and Fe.

\begin{table}[hp]
\begin{center}
\caption{Updated Settling Times and Accretion Rates for GD 40 and G241-6}
\begin{tabular}{lllcc}
\\
\hline \hline
	& t$_{set}$$^a$	& $\dot{M}$(Z)$_{GD 40}$	& $\dot{M}$(Z)$_{G241-6}$& \\
Z	& (10$^6$ yr)	& (g s$^{-1}$)	& (g s$^{-1}$) \\
\hline
C      & 1.1	& 2.2 $\times$ 10$^6$	&	$<$ 4.4 $\times$ 10$^5$\\
N      & 1.1	& $<$ 2.6 $\times$ 10$^5$	&	$<$ 2.1 $\times$ 10$^5$\\
O	& 1.1	& 4.5 $\times$ 10$^8$	&	4.3 $\times$ 10$^8$ \\
Mg	& 1.2	& 1.7 $\times$ 10$^8$	&	1.5 $\times$ 10$^8$ \\
Al	& 1.2	& 1.4 $\times$ 10$^7$	&	$<$ 6.1 $\times$ 10$^6$ \\
Si	& 1.0	& 1.3 $\times$ 10$^8$	&	8.7 $\times$ 10$^7$ \\
P      & 0.79	& 1.1 $\times$ 10$^6$	&	4.7 $\times$ 10$^5$ \\
S      & 0.64	& 1.0 $\times$ 10$^7$:	& 5.6 $\times$ 10$^7$	\\
Cl	& 0.51	& $<$ 8.0 $\times$ 10$^5$	& $<$ 5.8 $\times$ 10$^5$	\\
Ca	& 0.51	& 1.3 $\times$ 10$^8$	& 5.1 $\times$ 10$^7$ \\
Ti      & 0.49	& 3.2 $\times$ 10$^6$	& 1.4 $\times$ 10$^6$		\\
Cr	& 0.53	& 6.4 $\times$ 10$^6$	& 4.5 $\times$ 10$^6$	\\
Mn    & 0.53 	& 3.1 $\times$ 10$^6$	& 2.4 $\times$ 10$^6$	\\
Fe	& 0.56& 4.4 $\times$ 10$^8$	& 2.0 $\times$ 10$^8$	\\
Ni	& 0.61& 1.8 $\times$ 10$^7$	& 8.9 $\times$ 10$^6$	\\
Cu	& 0.58	& $<$ 1.8 $\times$ 10$^5$	& $<$ 1.8 $\times$ 10$^5$	\\
Ga	& 0.50	& $<$ 2.9 $\times$ 10$^4$	& $<$ 2.9 $\times$ 10$^4$	\\
Ge    & 0.43	& $<$ 1.4 $\times$ 10$^5$	& $<$ 1.4 $\times$ 10$^5$	\\
Total	&	& 1.4 $\times$ 10$^9$	& 9.9 $\times$ 10$^8$	\\
\hline
\end{tabular} \label{Tab: GD40G241-6}
\end{center}
$^a$ This column is for both GD 40 and G241-6 because their atmospheric conditions are similar.
\end{table}

\begin{figure}[hp]
\epsscale{1.1}
\plotone{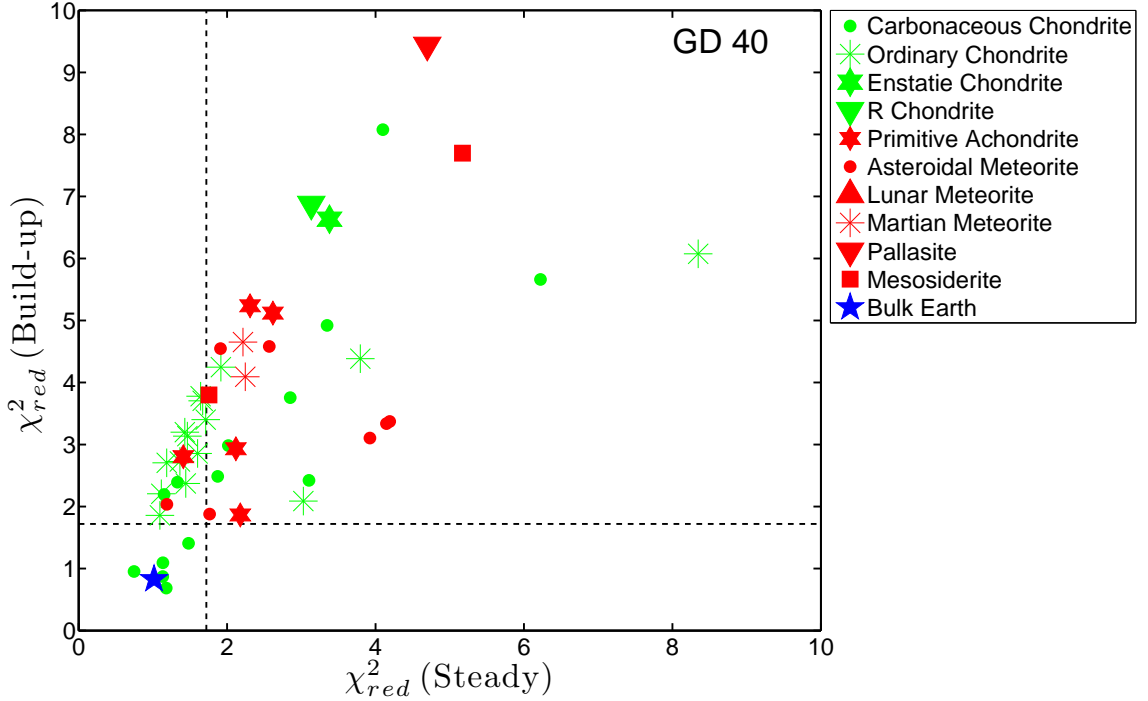}
\caption{Similar to Figure \ref{Fig: GD_chi} except for GD 40 comparing the mass fraction of 13 elements, including C, O, Mg, Al, Si, P, S, Ca, Ti, Cr, Mn, Fe and Ni, relative to the sum of Mg, Si, Fe and O. Both carbonaceous chondrites and bulk Earth are a good match to the parent body accreted onto GD 40.}
\label{Fig: GD40_chi}
\end{figure}

The total accretion rate for GD 40 is a factor of 2 lower than the value derived in \citet{Klein2010}, but the relative abundances change much less. The result of a $\chi^2_{red}$ analysis is presented in Figure \ref{Fig: GD40_chi}. When including all 13 detected elements,  both carbonaceous chondrites and bulk Earth can match the composition to 95\% confidence level for both steady-state and build-up approximations. The accreted planetesimal appears to be primitive and can be formed under nebular condensation, similar to what was concluded by \citet{Jura2012}. 

\begin{figure}[hp]
\epsscale{1.1}
\plotone{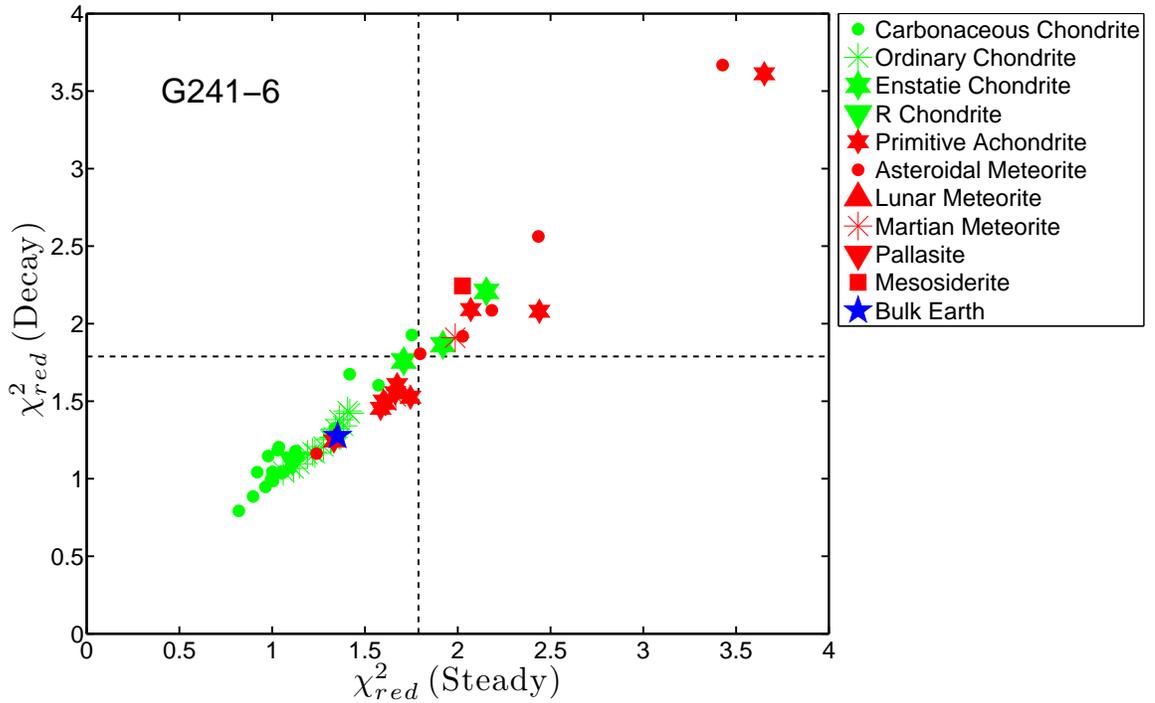}
\caption{Similar to Figure \ref{Fig: GD40_chi} except for G241-6 comparing 11 elements -- O, Mg, Si, P, S, Ca, Ti, Cr, Mn, Fe and Ni for steady-state versus decay phase when the accretion stopped 0.6 $\times$ 10$^6$ yr ago. Both carbonaceous chondrites and ordinary chondrites produce a good fit to the parent body accreted onto G241-6. Ordinary chondrites are a relatively better match because of the low carbon abundance, which is not considered in this $\chi^2_{red}$ plot because only an upper limit was reported for G241-6 \citep{Jura2012}.}
\label{Fig: G241-6_chi}
\end{figure}

The newly-derived total accretion rate for G241-6 is about a factor of 2 lower than previously reported \citep{Zuckerman2010}. The non-detection of an infrared excess and the slight depletion of heavier elements suggest that it may be at the beginning of a decay phase \citep{XuJura2012, Klein2011}. We assess both steady-state and decay phase for the $\chi^2_{red}$ analysis; in the latter case, we assume that accretion stopped 0.6 $\times$ 10$^6$ yr ago, approximately one settling time for Fe because its mass fraction is depleted by a factor of 2 relative to CI chondrites. The composition of the parent body is calculated following \citet{Zuckerman2011} and Equation (5) in \citet{Koester2009a}. A fuller exploration of different time-varying models will be presented in the future in the spirit of \citet{JuraXu2012}. As shown in Figure \ref{Fig: G241-6_chi}, both carbonaceous chondrites and ordinary chondrites provide good matches to all 11 elements, including O, Mg, Si, P, S, Ca, Ti, Cr, Mn, Fe and Ni. However, the carbon upper limit in G241-6, which is not included in the $\chi^2_{red}$ analysis, is at least one order of magnitude lower than most carbonaceous chondrites \citep{Jura2012}. Thus, ordinary chondrites are a more promising solar system analog to the parent body accreted onto G241-6 and nebular condensation is sufficient to produce the observed abundance pattern.

The $\chi^2_{red}$ analysis for GD 40 and G241-6 confirms the previous results \citep{Jura2012}; the accreted extrasolar planetesimals can be formed under nebular condensation and their compositions resemble primitive chondrites in the solar system.

\end{CJK}

\bibliographystyle{apj}
\bibliography{apj-jour,Ref.bib}

\end{document}